\documentclass[journal,10pt]{IEEEtran}
\usepackage[pdfpagelabels,
           plainpages=false,
           bookmarks=true,
           hypertexnames=false,
           pdffitwindow=false,
           colorlinks=true,
           linkcolor=blue,
           citecolor=blue,
           filecolor=black,
           urlcolor=black,
           pdfauthor={Mohamed Ahmed, Michele Garetto, Paolo Giaccone, Emilio Leonardi, Saverio Niccolini and Stefano Traverso},
           pdftitle={Unravelling the Impact of Temporal and Geographical Locality in Today's Content Caching Systems},
           pdfsubject={2014},
           pdfkeywords={Caching, LRU},
           pdfproducer={Latex with hyperref},
           pdfcreator={pdflatex}
          ]{hyperref}

\usepackage{epstopdf}
\usepackage{color}
\usepackage{cancel}
\usepackage{xspace,epsfig}
\usepackage{graphicx,times,latexsym,amsfonts,amssymb,amsmath}
\usepackage{comment,url,color}
\usepackage{multirow}
\usepackage{caption}
\usepackage{subcaption}
\usepackage{enumitem}

\def\newblock{\relax}

\newcommand{\equaref}[1]{(\ref{eq:#1})}

\newcommand{\Uno}{ \mathbb{I} }

\newcommand{\diff}{{\rm\,d}}
\newcommand{\Prob} { {\rm Pr}}
\newcommand{\lambdahat}{\overline{\lambda}}

\newtheorem{teorema}{\bf Theorem}
\newtheorem{corollario}{\bf Corollary}

\newenvironment{remark}%
{ {\bf Remark.~~}} %
{\par\noindent} %

\newcommand{\ls}[1]
{\dimen0=\fontdimen6\the\font
  \lineskip=#1\dimen0
  \advance\lineskip.5\fontdimen5\the\font
  \advance\lineskip-\dimen0
  \lineskiplimit=.9\lineskip
  \baselineskip=\lineskip
  \advance\baselineskip\dimen0
  \normallineskip\lineskip
  \normallineskiplimit\lineskiplimit
  \normalbaselineskip\baselineskip
  \ignorespaces
}

\newcommand{\tgifeps}[3]{
  \begin{figure}[t]
    \centering
    \includegraphics[width=#1cm, clip=true]{#2}
    \vspace{-0.1cm}
    \caption{#3\label{fig:#2}}
  \end{figure}

}

\includecomment{conference}
\excludecomment{techrep}
\excludecomment{conference}
\includecomment{techrep}

\newcommand{\mmm}{\vspace*{-1mm}}

\newcommand{\be}{\begin{equation}}
\newcommand{\ee}{\end{equation}}
\newcommand{\ba}{\begin{array}}
\newcommand{\ea}{\end{array}}

\newcommand{\RMED}[2]{}

\newcommand{\FWPDF}{\textit{Home~1}}
\newcommand{\FWPUL}{\textit{Home~2}}
\newcommand{\FWPULPDF}{\textit{Home~3}}
\newcommand{\FWUMB}{\textit{Home~4}}
\newcommand{\POLI}{\textit{Campus~1}}
\newcommand{\TKPL}{\textit{Home~5}}

\newcommand{\TONE}{\textit{Trace~1}\xspace}
\newcommand{\TTWO}{\textit{Trace~2}\xspace}
\newcommand{\TTHREE}{\textit{Trace~3}\xspace}
\newcommand{\TFOUR}{\textit{Trace~4}\xspace}
\newcommand{\TFIVE}{\textit{Trace~5}\xspace}
\newcommand{\TSIX}{\textit{Trace~6}\xspace}
\newcommand{\TSEVEN}{\textit{Trace~7}\xspace}
\newcommand{\TEIGHT}{\textit{Trace~8}\xspace}
\newcommand{\TNINE}{\textit{Trace~9}\xspace}
\newcommand{\TTEN}{\textit{Trace~10}\xspace}

\title{Unravelling the Impact of Temporal and Geographical Locality in Content Caching Systems}

\author{
\begin{center}
 Stefano Traverso, Mohamed Ahmed,  Michele Garetto,\\
 Paolo Giaccone,  Emilio Leonardi, Saverio Niccolini
\end{center}
\thanks{S.\ Traverso, P.\ Giaccone, E.\ Leonardi are with the Department of Electronic and Telecommunications, Politecnico di Torino, Italy: 
e-mail: \{lastname\}@tlc.polito.it. 
M.\ Ahmed, S.\ Niccolini are with the NEC Labs Europe, Heidelberg, Germany:
e-mail: \{name.lastname\}@neclab.eu. 
Michele Garetto is with Universit\`a di Torino, Italy;
e-mail: michele.garetto@unito.it.}\vspace{-7mm}
}

\begin{document}

\maketitle
\begin{sloppypar}
\begin{abstract}
To assess the performance of caching systems, the definition of a  
proper process describing the content requests generated by users is required.
Starting from the analysis of traces of YouTube video requests  
collected inside operational networks, we identify
the characteristics of real traffic that need to be 
represented and those that instead can be safely neglected.
Based on our observations, we  introduce a simple, parsimonious traffic  model, 
named Shot Noise Model (SNM), that allows us to 
capture temporal and geographical locality of content popularity.   
The SNM is sufficiently simple to be effectively employed in both analytical and
scalable simulative studies of caching systems.
We demonstrate this by analytically characterizing the 
performance of the LRU caching policy under the
SNM, for both a single cache and a network of caches. 
With respect to the standard Independent Reference Model (IRM),
some paradigmatic shifts, concerning the impact of various traffic 
characteristics on cache performance,
clearly emerge from our results.
\end{abstract}

\mmm\mmm
\section{Introduction}\label{sec:introduction}

It is no surprise to find that the design and analysis of content caching
systems continue to receive attention from both industry and
academia. The big players in the market (Google, Akamai,
Limelight, Level3, etc.), today preside over a multi-billion dollar
business built on content delivery networks (CDNs), which employ
massively distributed networks of caches to carry over half of Internet 
traffic, according to recent measurements~\cite{ciscoVNI}. 
To illustrate this reality, in 2010 Akamai listed
its CDN to include over 60,000 servers in 1000 networks, spread over 
70 countries~\cite{Nygren:2010}. 
The impressive growth of CDNs is essentially 
driven by the explosion of multimedia traffic.
It is expected that video traffic alone will be around
70 percent of all consumer Internet traffic in 2017, 
and almost two-thirds of it will be delivered by CDNs~\cite{ciscoVNI}.

The fundamental role played by caching systems goes beyond existing
CDNs.  Indeed, a radical change of communication paradigm
may take place in the  future Internet, from the traditional host-to-host
communication model created in the 1970s, to a new host-to-content
kind of interaction, in which the main networking functionalities
are directly driven by object identifiers, rather than host
addresses. In particular,
Content-Centric-Networking proposals (CCN)~\cite{Jacobson_icn} aim at
redesigning the entire global network architecture with named data as
the central element of the communication. 
This translates in practice into the
need to redesign core routers by equipping them with fast, small
caches, capable of 
processing requests at line speed.   
Nonetheless, to date, the design and evaluation of large-scale,
interconnected systems of caches is still poorly understood.  
In the first place, it is unclear how to properly describe
the traffic (in terms of the sequence of contents' requests) 
generated by the users, that is then processed by cache networks. 
In this regard, just resorting to trace-driven simulations to assess the performance
of a cache architecture clearly has severe limitations, as we will 
elaborate in the next section.
It is therefore highly desirable to have, first of all, 
a proper model for the arrival process of contents' requests at
the caches.  The main challenge here is to find a good compromise
between: i) the fidelity of the model in describing the behavior of real traffic;
and ii) its simplicity, which permits the development of analytical 
tools to predict the system performance. 
To fully address this problem, one needs to identify the
traffic features playing the most crucial role for the resulting 
cache performance, and capture them into in a flexible,
parsimonious, and analytically tractable manner. 
To the best of our knowledge, this problem has not yet received a satisfactory
answer.

\vspace{-3mm}
\subsection{The necessity of a  traffic model} 
\label{sec:necessity}

Although caching systems continue to attract interest in the
networking community, there is still no common agreement on the
traffic assumptions under which system design and performance
evaluation should be carried out, especially in the context of
pervasive CDN and CCN architectures.  To obtain the best fidelity in
evaluating a given system, one could simply follow the approach of
performing trace-driven analysis
~\cite{Jiang_conext12,Poese:2012}. 
This approach, however, has several
shortcomings.  First, it does not enable us to identify the important
factors that influence system performance and understand their role.
Second, it does not permit us to explore ``what if'' scenarios, such
as: how will my system perform if I increase/modify the catalogue of
available contents? or if the users' population becomes much larger?
Third, too often, we are constrained by the size and/or the
availability of data sets, their diversity as well as legal and
privacy concerns, which impacts the fidelity of the analysis.

To overcome these limitations, we can instead analyze caching systems
using synthetic traces produced by a traffic model.  The simplest, and
still most widely adopted traffic model in the cache literature~\cite{Recent_IRM1,Gallo,Roberts_mix,Roberts-ITC} is the
so-called Independent Reference Model (IRM)~\cite{Coffman:73},
according to which the sequence of content requests arriving at a
cache is characterized by the following fundamental assumptions: i)
there exists a fixed catalogue of $N$ distinct contents, which does not
change over time; ii) the probability a request is for a specific
content 
is {\em constant} (i.e.\ the content popularity does
not vary) and {\em independent} of all past requests.  The IRM is
commonly used in combination with a Zipf-like law of content
popularity.  In its simplest form, Zipf's law states that the
probability to request the $n$th most popular content is proportional to
$1/n^\alpha$, where the exponent $\alpha$ depends on the considered
system (especially on the type of contents)~\cite{Roberts_mix}, and
plays a crucial role on the resulting cache performance.

By construction, the IRM completely ignores all temporal correlations
in the sequence of requests. In particular, it does not take into
account a key feature of real traffic, usually referred to as {\em
  temporal locality}, which occurs when requests for a given content
densify over short periods of time (with respect to the trace
duration).  The important role played by temporal locality, especially
its beneficial effect on cache performance, is well known in the
context of computer memory architecture~\cite{Coffman:73} and web
traffic~\cite{Fonseca:03}.  Indeed, several extensions of IRM have
been already proposed to incorporate temporal correlations in the
request process~\cite{Coffman:73,Fonseca:03,Crovella,Bestavros}, 
i.e., the fact that, if a content is
requested at a given time instant, then it is more likely that the
same content will be requested again in the near future.  Existing
models, however, have been primarily thought for web traffic, and they
share the following two assumptions: i) the content catalogue is
fixed; ii) the request process for each content is stationary (i.e.,
either a renewal process or a semi-Markov-modulated Poisson
process or a self-similar process). 
As we will see, these assumptions are not
appropriate to capture the kind of temporal locality usually
encountered in Video-on-Demand traffic,
because 
they do not easily capture 
intrinsically non-stationary macroscopic effects related to content popularity dynamics.
Moreover some of the previously proposed models are too complex to allow an analytical study of caching systems.

At the other extreme, some recent studies have proposed rather
sophisticated models describing the evolution of contents popularity
at the macroscopic level. These show that the aggregate download
process of popular on-line contents is highly non-stationary and
exhibits a complex correlation structure resulting from social
cascades and other viral phenomena~\cite{Crane2008,mat2012}.  Fairly
complex stochastic models, i.e.\ based on Hawkes
processes~\cite{Crane2008} or autoregressive (ARIMA)
models~\cite{mat2012}, have been recently proposed to accurately
describe the large-scale content popularity evolution.  However, it is
unclear how these models can be used to generate a synthetic sequence
of content requests arriving at a specific cache, which aggregates
traffic from a limited number of users.

Furthermore, with respect to distributed systems of caches, the {\em
  geographical locality} of user requests has been little investigated in
the literature and largely ignored by existing analytical
models. Large-scale, pervasive systems of caches typically serve
heterogeneous communities of users having different interests, and
therefore the probability of a request for a given content can vary
significantly from region to region. The studies that have
appeared~\cite{Scellato:11,Brodersen12,Huang:2013}, show that geographical
locality is (as expected) hardly observable within culturally
homogeneous regions, and becomes evident in large systems
serving different linguistic/cultural communities.  Our results (see
Sec.~\ref{sec:spatial}) show that geographical locality can be observed to
some extent even in limited geographical regions because users
associated to different caches vary in their social/ethnic/linguistic
composition.

To the best of our knowledge, no traffic models have been proposed so
far to incorporate various degrees of geographical locality in the content
request processes arriving at different caches.  Furthermore, the
impact of geographical locality on cache performance, especially in
distributed systems of interconnected caches, is still poorly
understood~\cite{Borst,Applegate}. 

\vspace*{-0.3cm}
\subsection{Paper contributions}

Because of its dominant and growing role in the Internet, 
in this paper we focus on video traffic, specifically, Video-on-Demand (VoD). 
Nonetheless, our methodology and results
have broader applicability, and may also be of interest to other kinds
of contents.  We are especially interested in pervasive CDN
or CCN architectures, comprising several caches (which can be small
relative to the content catalogue size) serving localized communities
of users. 
However, in this work, we do not consider the case in which users are spread 
over many different time zones.

Our first main contribution is a new traffic model describing the
contents' request processes originated by the users, to be used in
input to the edge caches of the system. In pursuing this goal, we
aimed at filling the
existing gap between simple, stationary traffic generators developed
for traditional caching systems (computer architecture, web
traffic)~\cite{Coffman:73,Fonseca:03} and the complex stochastic
models describing macroscopic world-wide content popularity
dynamics~\cite{Crane2008,mat2012}.  Our proposed traffic model
meets the following requirements: (1) be general and flexible; (2)
provide a native explanation for the temporal and geographical locality in
the request process; (3) explicitly represent content popularity
dynamics; (4) capture the phenomena having major impact on cache
performance, while neglecting those with no or limited impact; (5) be
as simple as possible while maintaining accuracy; and (6) permit
developing analytical models of popular caching policies.

Our second main contribution is an accurate analysis of the LRU (Least
Recently Used) caching policy under the newly proposed traffic model,
that provides fundamental insights on the impact of several traffic
characteristics on cache performance, which have not been documented
before.

In more detail, we provide the following contributions, listed in the
sequence in which they are derived in the paper:
{\bf(C1)} We analyze the temporal and geographical locality of
real Video-on-Demand traffic collected in six different locations (Sec.~\ref{sec:analysis}).
{\bf(C2)} We show that the standard IRM approach to modeling users' contents requests
leads to significant errors when estimating the cache size needed to achieve a given hit
ratio, especially when cache sizes are small relative to the catalogue
size (Sec.~\ref{sec:temporal_locality}).  {\bf(C3)} We propose and
validate (using our traces) the Shot Noise Model (SNM), a more
accurate and flexible model alternative to the IRM
(Sec.~\ref{sec:traffic_model}).  {\bf(C4)} We show that, in contrast
to common expectation, daily variations in the aggregate request rate, as well as
the detailed shape of the popularity profile of individual contents, have negligible impact
on cache performance, advocating the idea of a parsimonious traffic model
(in terms of the number of parameters) (Sec.~\ref{subsec:valid_single}).
{\bf(C5)} We explain how the SNM can be extended to the case of a cache
network, validating some simplifying assumptions that we propose
to incorporate geographical locality in the model 
(Sec.~\ref{subsec:snm_multi} and~\ref{subsec:validate_multi}).
{\bf(C6)} We {\em analytically} characterize the performance of the LRU caching policy under the SNM,
providing enlightening closed-form expressions for the large and small cache
regimes (Sec.~\ref{sec:modelling}).  {\bf(C7)} Using
numerical and simulative analysis, we explore the effect
of a wider range of model parameterizations than the one available from the
traces (Sec.~\ref{sec:validation}), gaining deeper insights into the
impact of  several traffic parameters, which in some cases
depart from the general understanding gained using the IRM.
(Sec.~\ref{sec:singleclass}). {\bf(C8)}
At last, we show how our traffic model and analysis
can be effectively used for system design and optimization,
investigating some additional examples of traffic mixes
(Sec.~\ref{sec:multiclass} and \ref{sect:result-multicache}).
 
\mmm\mmm
\section{A Dive into Real  Traffic} \label{sec:analysis}

\begin{table}[b]
\tiny
\centering
\begin{tabular}{|c|c|c||c|c|c|}
\hline
Probe   & Type        & IPs   & Probe       & Type    & IPs\\
\hline
\FWPDF\   & ADSL        & 16172   & \FWUMB\   & ADSL    & 2543  \cr
\FWPUL\   & ADSL/FTTH     & 17242   & \TKPL\    & ADSL    & 5080  \cr
\FWPULPDF\  & ADSL/FTTH     & 31124   & \POLI\    & LAN/Wi-Fi  & 15000  \cr
\hline
\end{tabular}
\caption{Probe characteristics.}
\label{tab:desc-pops}
\vspace*{-0.1cm}
\end{table}

\begin{table}[b!]
\tiny
\centering
\begin{tabular}{|c|c|c|c|c|c|c|}\hline
Probe                       & Trace     & Period    & Length  & Requests & Videos \cr\hline\hline
\multirow{3}{*}{\FWPDF}     & \TONE\    & 20/03/12-25/04/12   & 35 days & 1.7M     & 0.93M\cr
                            & \TTWO\    & 30/04/12-28/05/12   & 27 days & 1.8M     & 0.95M\cr
                            & \TSIX\    & 18/03/13-24/04/13   & 36 days & 1.6M     & 0.86M\cr\hline
\multirow{2}{*}{\FWPUL}     & \TTHREE\  & 20/03/12-30/04/12   & 40 days & 2.4M     & 1.24M\cr
                            & \TSEVEN\  & 12/03/13-24/04/13   & 42 days & 2.4M     & 1.22M\cr\hline
\multirow{2}{*}{\FWPULPDF}  & \TFOUR\   & 20/03/12-25/04/12   & 35 days & 3.8M     & 1.76M\cr
                            & \TTEN\    & 18/03/13-24/04/13   & 36 days & 3.7M     & 1.69M\cr\hline
\FWUMB\                     & \TFIVE\   & 15/02/13-23/04/13   & 67 days & 0.4M     & 0.25M\cr\hline
\TKPL\                      & \TEIGHT\  & 28/02/13-21/03/13   & 21 days & 0.6M     & 0.28M\cr\hline
\POLI\                      & \TNINE\   & 7/03/12-13/05/12    & 67 days & 0.55M    & 0.35M\cr
\hline
\end{tabular}
\caption{Measurement traces.}
\label{tab:desc-traces}
\vspace*{-0.2cm}
\end{table}

We employed  Tstat\footnote{\url{http://www.tstat.polito.it}} an
open-source traffic monitoring tool to
analyze TCP/IP packets sent/received by actual end-users, captured at
monitored vantage points.  Probes were installed in five PoPs located
at different cities of two different countries, Italy and
Poland.  Table~\ref{tab:desc-pops} provides details about our
vantage points inside the network.
Probes \FWPDF, \FWPUL, \FWPULPDF\ and \FWUMB\ are located in three
cities in Italy, and monitor the traffic of about 65,000
residential customers of a large ISP offering Internet access by ADSL
and FTTH technologies.  Similarly, \TKPL, located in Poland,
monitors the network activity of approximately 5,000 residential
customers.  Finally, probe \POLI\ was
deployed within the network backbone of Politecnico di Torino in Italy, which provides
Internet access to about 15,000 students mainly through Wi-Fi access.

\begin{figure}[t!]
  \begin{minipage}[t]{0.48\columnwidth}
    \hspace*{-0.4cm}
    \includegraphics[width=1.15\linewidth]{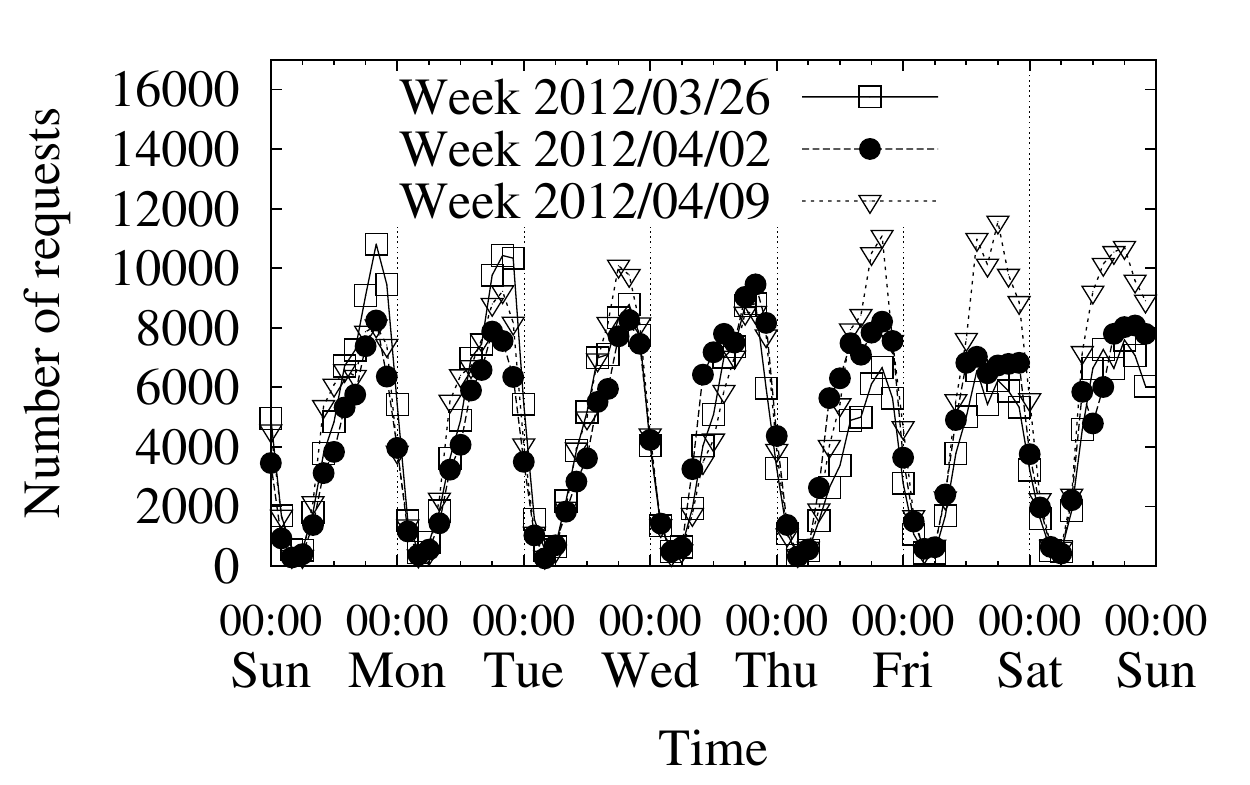}
    \caption{Evolution of the volume of requests over three weeks for \TTHREE.}
    \label{fig:reqs_vs_time_week_profile_trace_03}
  \end{minipage}
  \hspace*{0.1cm}
  \begin{minipage}[t]{0.48\columnwidth}
    \hspace*{-0.3cm}
    \includegraphics[width=1.15\textwidth]{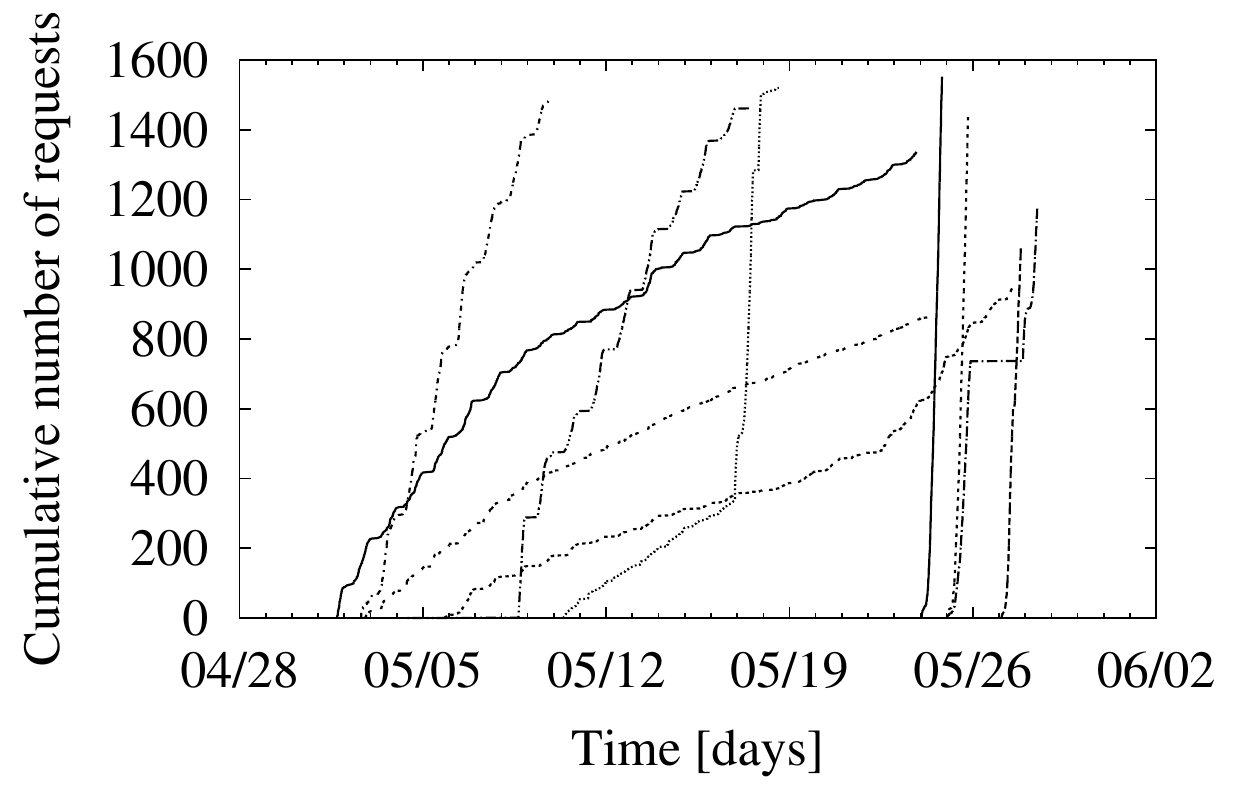}
    \caption{Cumulative number of requests over time for a subset of
videos in {\TTWO}.}
    \label{fig:reqs_vs_lifetime_trace_02}
  \end{minipage}
  \vspace*{-0.4cm}
\end{figure}

Table~\ref{tab:desc-traces} details the ten traces employed in our
study. Measurements were performed on both incoming and outgoing
traffic over two different periods (March-May 2012 and February-April
2013) and together cover approximately 6 months. In total, we observed
the activity of about 85,000 end-users accessing the Internet, and
identified the TCP flows corresponding to YouTube video requests and
downloads.  
In total, we recorded more
than 20 million transactions.

\mmm\mmm
\subsection{Temporal locality} \label{sec:temporal_locality}
From analyzing the traces, we observe two main factors responsible
for the temporal locality in the sequence of requests made by
users. First, the {\em aggregate} arrival rate of requests follows the
expected diurnal variation, as shown in
Fig.~\ref{fig:reqs_vs_time_week_profile_trace_03}.  Second, the
arrival rate of requests for a given content can be highly
non-stationary, being often concentrated in intervals much shorter
than the total trace duration. Furthermore, as illustrated in
Fig.~\ref{fig:reqs_vs_lifetime_trace_02}, contents display a wide
range of popularity evolution patterns~\cite{abrahamsson_imc2012}.
Although these facts are well known and have been already examined
before, their impact on cache performance must
be carefully evaluated.

With regard to diurnal variations, we observe that, contrary to
what is sometimes believed~\cite{abrahamsson_imc2012}, accounting
for this variation has no impact on the main performance metrics for
caches such as the hit probability (see Sec.~\ref{sec:diurnal}).
To intuitively understand why this is the case, consider that the hit
probability of almost all proposed caching policies depends only on
the sequence of content identifiers arriving at the cache, and not on
the time-stamps associated with the requests. Therefore, if we
arbitrarily squeeze or stretch (over time) the aggregated sequence of
content requests arriving at a cache,
we obtain the same hit probability (this holds for all caching
policies which do not explicitly use the information about the request
arrival time). For this reason, in our synthetic traffic model we will 
ignore the diurnal rate variations.

On the other hand, the non-stationarity of the popularity of
individual contents has a significant impact on the performance of a
cache. Accounting for this property is a difficult task due to the
complexity and heterogeneity of content popularity 
dynamics.  For example, the popularity of some videos vanishes after
only a few days, while others continue to attract requests for
significantly long periods of time (months or even years).
Besides the life-span, clearly also the number of
requests attracted by the videos can be very diverse.

\tgifeps{5.5}{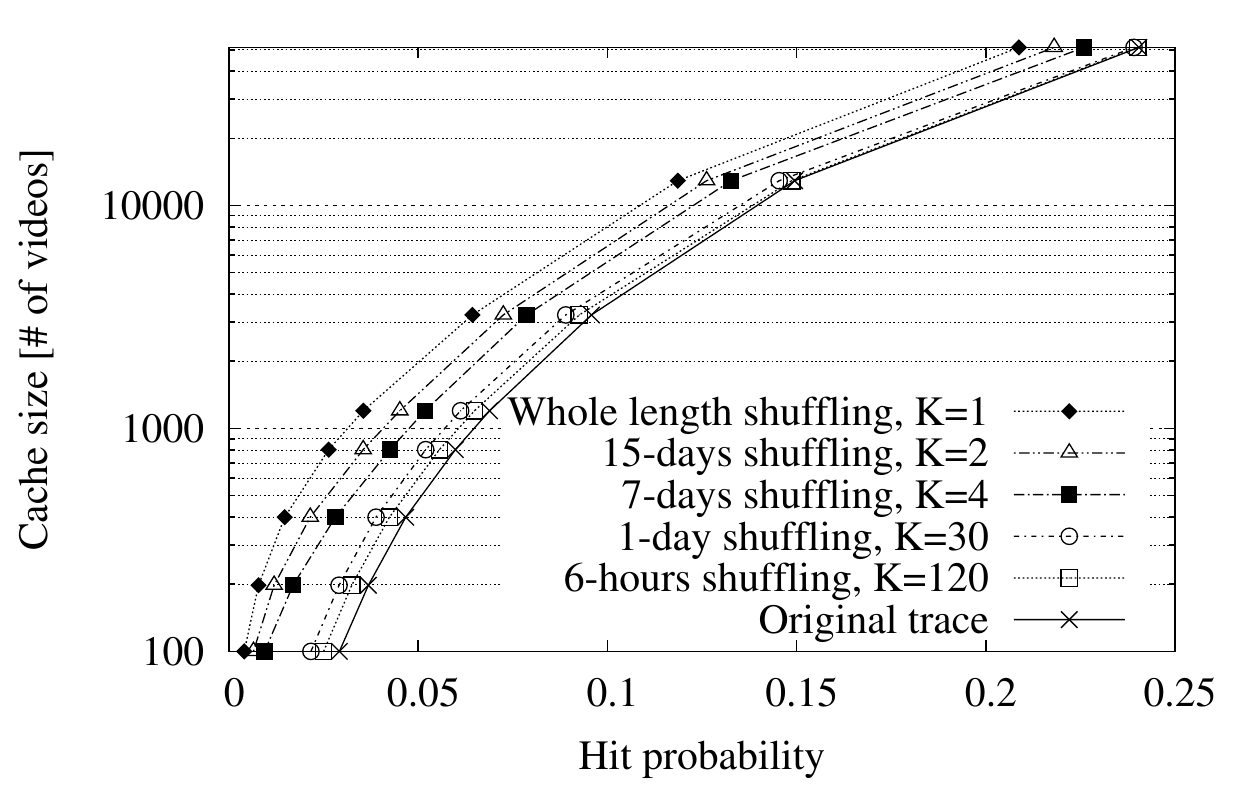}{The cache size
  required to achieve a desired hit probability, when an LRU cache is
  fed by the requests contained in \TONE, subject to different degrees
  of trace reshuffling. \vspace*{-0.3cm}}

\begin{figure*}[t!]
  \centering
  \hspace*{-0.1cm}
  \begin{subfigure}[t]{0.315\textwidth}
    \centering
    \includegraphics[width=1.05\linewidth]{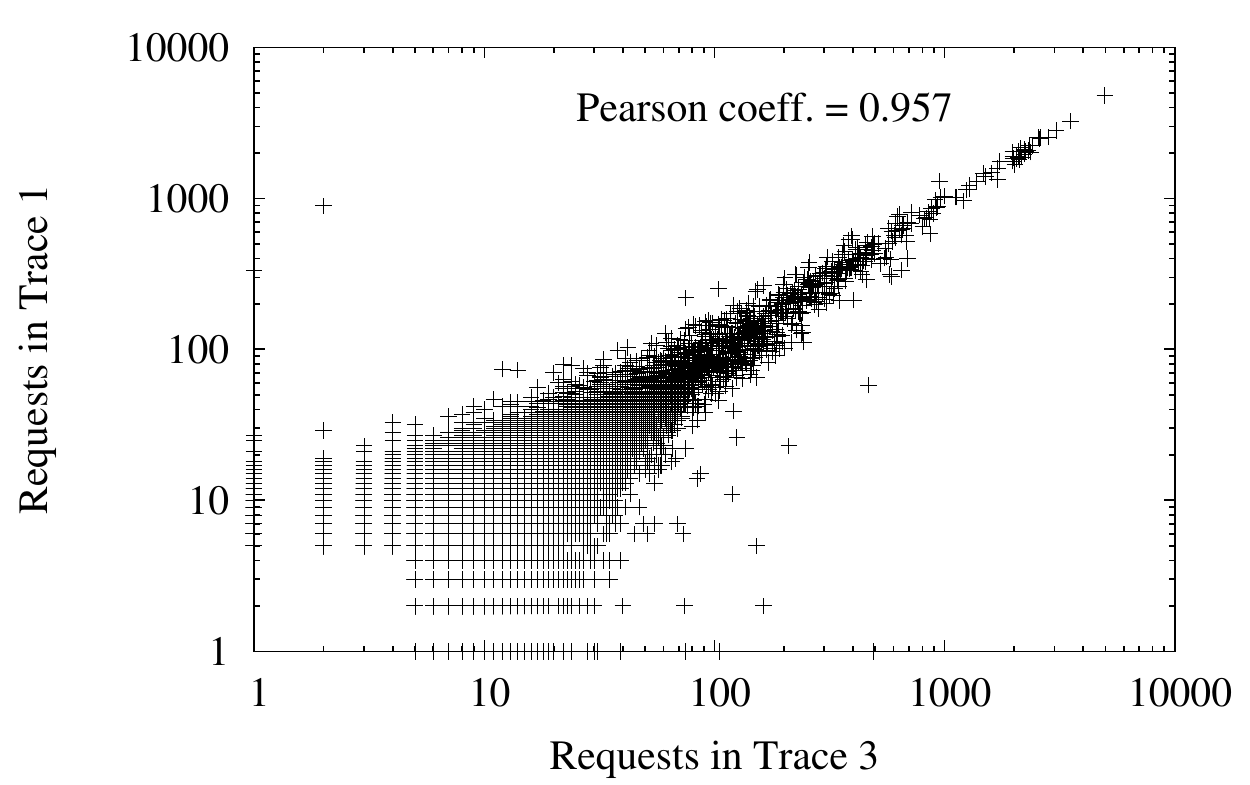}
    \caption{{\it Trace~1} and {\it Trace~3} (same country).}
    \label{fig:spatial_corr_high}
  \end{subfigure}%
  \hspace*{0.1cm}
  \begin{subfigure}[t]{0.315\textwidth}
    \centering
    \includegraphics[width=1.05\linewidth]{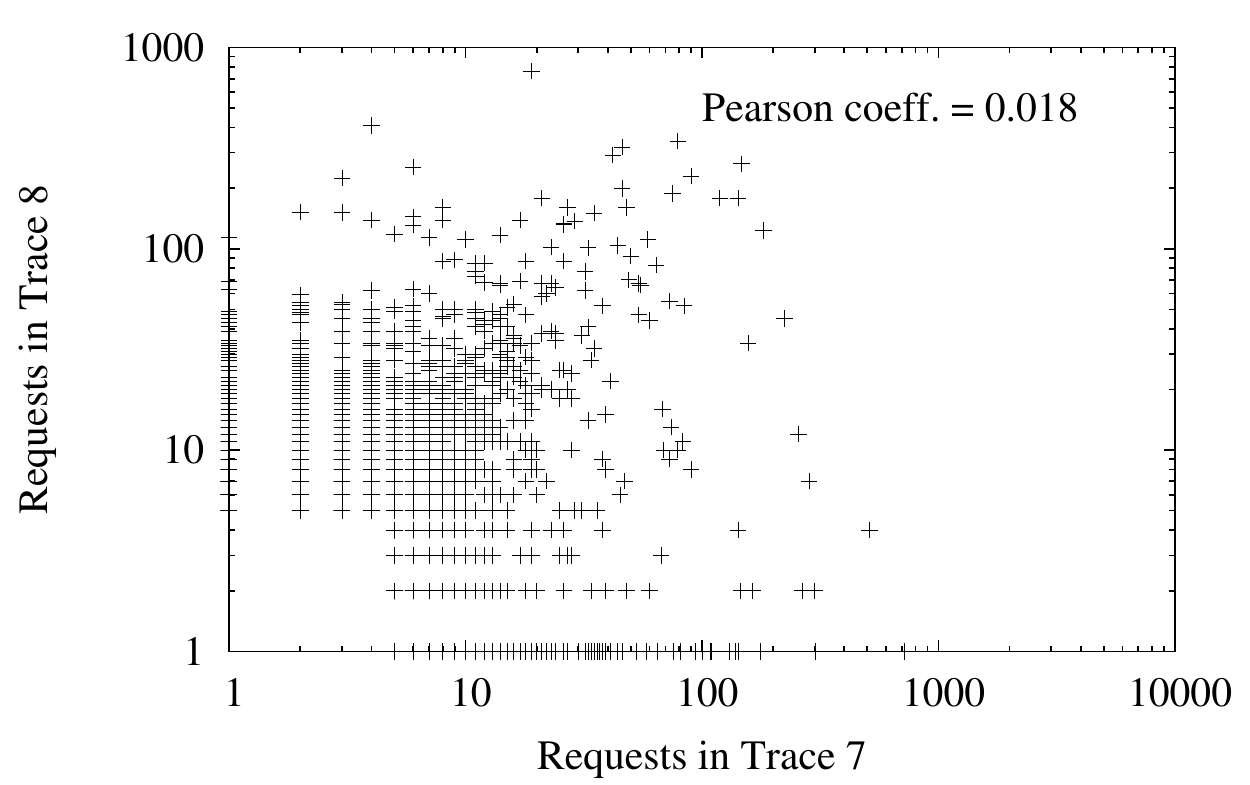}
    \caption{{\it Trace~7} and {\it Trace~8} (different countries).}
       \label{fig:spatial_corr_low}
  \end{subfigure}
  \hspace*{0.1cm}
  \begin{subfigure}[t]{0.315\textwidth}
    \centering
    \includegraphics[width=1.05\linewidth]{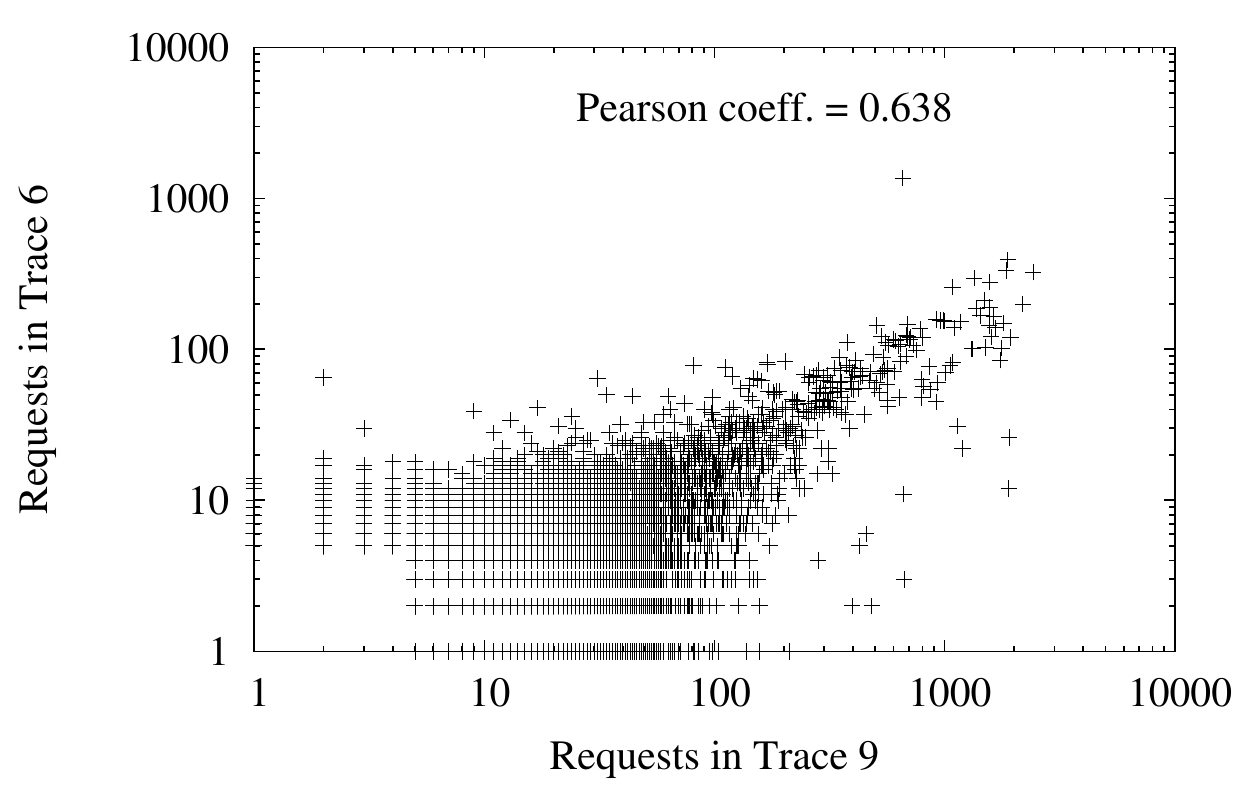}
    \caption{{\it Trace~6} and {\it Trace~9} (same country).}
       \label{fig:spatial_corr_mid}
  \end{subfigure}
  \vspace*{-0.1cm}
   \caption{Scatter-plot of the number of content requests in trace pairs. Each point corresponds to a specific content.}
  \vspace*{-0.3cm}
\end{figure*}
To better understand the impact on cache performance of the temporal locality 
present in our traces, we carried out the following
experiment: we fed an LRU cache (initially empty, and transient effects have been 
filtered out in the evaluation) with
the sequence of requests contained in \TONE~(similar results were obtained
using the other traces) and derived the cache
size necessary to achieve a given cache hit probability.  Results are
reported in \autoref{fig:phit_vs_cachesize_shuffle_flip_trace_01} by
the curve labeled ``Original trace''.
Then, we partitioned \TONE~into $K$ slices, each
containing an equal number of requests, and we randomly
permuted the requests within each slice. Such artificially shuffled
traces (one for each $K$) are then fed again into an LRU cache,
deriving again the required cache size to achieve a given hit probability.
Different values of $K$ correspond to
washing out the temporal locality at a time scale equal
to the corresponding slice duration (for clarity, the approximate
slice duration is also reported in the legend of \autoref{fig:phit_vs_cachesize_shuffle_flip_trace_01},
for each considered value of $K$).
Note that the original trace can be considered
as a limit case of very large $K$ (equal to the number of requests in the trace),
whereas $K = 1$ corresponds to the case in which we randomly permute
the entire trace, destroying all temporal correlations within the whole trace duration (about one month).

As expected, we find that temporal locality plays a significant
role on cache performance -- the cache size needed to achieve a desired
hit probability increases considerably between the original trace and
the extreme case with $K=1$.  
Now, suppose that we were to completely ignore the effects of
temporal locality, by adopting a naive IRM approach in which we just
compute from the trace the empirical popularity distribution for the
contents requested in the trace, and use such empirical distribution
as the popularity law of the IRM. The hit probability achieved for a
given cache size, according to the above IRM model, would be
essentially equivalent to the one derived in our experiment in the
case $K=1$.  Indeed, the complete trace shuffling leads to an
i.i.d.\ sequence of requests following the long-term empirical
probability distribution of the trace, just as in the considered IRM
model. In conclusion, the adoption of a naive IRM
approach for VoD contents leads to considerably erroneous
(pessimistic) estimates of cache performance, 
especially when caches are small.

We also observe that, as the slice duration approaches the order of a few hours, the
required cache size becomes very close to the one resulting from
the original trace. This means that: i) the evolution of content
popularity over timescales of few days/weeks is important for the
resulting cache performance; ii) we can, instead, ignore short
timescale effects (i.e., correlations taking place over timescales
of few hours or less).  The practical consequence of this is that we
do not need to take into account complex fine-grained correlations
in the arrival process of requests (in particular at the level of
contents' inter-request times). Actually, given that short
time-scale correlations (up to a few hours) are not important to
predict cache performance for VoD contents, we can well adopt
(locally) a Poisson approximation for the arrival process of
requests, which enables us to build simple analytical models 
of cache behavior, such as those developed in
Sec.~\ref{sec:modelling}.

We remark that our findings are in sharp contrast with 
what has been observed in the case of web traffic, whereby
the impact of short time-scale correlations greatly outweighs that due to 
long-term correlations~\cite{Mahanti}. This observation signifies the
different nature of VoD with respect to web browsing, 
motivating the development of new models specifically tailored to 
this kind of traffic.

We emphasize that the IRM approach could, in principle, still be adopted
to effectively predict the hit probability in the scenario
considered in~\autoref{fig:phit_vs_cachesize_shuffle_flip_trace_01}. This would require
us to estimate from the trace a proper content catalogue size (which is a non-trivial task in its own) 
and to properly set the contents popularity law of IRM so as to match the short-term
(say, over a few hours) distribution of content popularity observed in the trace.
Estimating a short-term popularity distribution from a trace, however,
is a challenging task (as recognized in~\cite{Roberts-ITC}), being
any measurement collected over a short period necessarily affected by a large
amount of noise. Moreover, the proper timescale at which the IRM could
be effectively employed is hard to predict, as it depends on many
parameters (cache size, caching policy, arrival rate of requests, etc.), forcing us
to compute a different short-term popularity distribution for each
considered scenario. On the contrary, the parameters of our traffic model
can be derived, once for all, from long-term measurements, and our analysis
of cache performance does not require to explicitly compute
any short-term popularity law.

\mmm\mmm\mmm\mmm
\subsection{Geographical locality} \label{sec:spatial}
Geographical diversity in the contents' popularity
is expected to play a significant role in a large-scale systems of interconnected caches,
in which edge caches receive the requests of subsets of users geographically
localized in the same region, and thus likely to share similar interests.
Thanks to the significant time-overlap between some 
of the traces belonging to our data set (see Table~\ref{tab:desc-traces}),
we were able to assess, to some extent, the geographical locality of VoD traffic.  
In particular, we counted the number of requests received by each video,
during the largest common time interval between two given traces of our
data set. Fig.s~\ref{fig:spatial_corr_high}-\ref{fig:spatial_corr_mid}
show the resulting number of requests in one trace vs.\ the one in
the other trace, for three significant cases.  Note that a perfect
linear relation here would imply exactly the same relative popularity of 
contents in the two different traces, corresponding to 
a Pearson correlation coefficient equal to 1.
In Fig.~\ref{fig:spatial_corr_high} we consider two traces collected 
at probes belonging to the same residential ISP in one country (Italy), 
i.e., serving highly homogeneous users in terms of
language and culture.  As expected, contents which are very popular in
one trace are also very popular in the other trace, as confirmed by the
Pearson coefficient very close to 1.  In contrast,
Fig.~\ref{fig:spatial_corr_low} considers two traces collected at
probes located in different countries (Italy and Poland), whose native languages are different. In this case, we observed a very low correlation (Pearson coefficient close to 0), suggesting that the cultural background of
users plays a crucial role in the
diversity of the contents request process\footnote{We verified that the few
YouTube videos attracting a large volume of requests in both traces
considered in Fig.~\ref{fig:spatial_corr_low} are related to famous international hits.}.

Although country borders may represent a good baseline to identify
clusters of homogeneous users, we observed that even within the same
country there can be significant heterogeneity in terms of users'
interests.  In Fig.~\ref{fig:spatial_corr_mid} we
compare a trace collected in a residential network (PoP \FWPDF) with
the trace from a university campus network (\POLI), on a time-overlap of about 
one month. Even though both traces are collected in the same city, 
a significantly higher fraction of points lie far from the diagonal,
with respect to the scenario in Fig.~\ref{fig:spatial_corr_high},
especially in the case of popular videos. This is confirmed 
by the correlation coefficient, here equal of $0.638$. 

We conclude that accounting for geographical locality in a traffic model
can be a difficult task due to cultural/social effects. A deep
assessment of the amount of geographical diversity in the network, by
means of large measurement campaigns, would be required to 
build accurate synthetic traffic models. Unfortunately, due to the limited
available data set, we were able to investigate the impact of geographical
locality on cache performance only based on synthetic traffic
traces, as discussed later in Sec.~\ref{sect:result-multicache}.

\vspace*{-0.1cm}

\section{Shot noise traffic model}\label{sec:traffic_model}
Guided by the insights gained from our traces (Sec.~\ref{sec:analysis}), 
we propose a new traffic model, aimed at striking a good compromise
between simplicity, flexibility and accuracy.
We first consider the single cache case and then move on to the case of a network
of caches.

\vspace*{-2mm}
\subsection{Basic model for a single cache}\label{subsec:shotonecache}
The rationale of our traffic model is to capture the physical origin 
of the temporal locality observed in the traces.
Our solution is to represent the overall request process as
the superposition of many independent processes, each referring to an
individual content. As such,
the arrival process of a given content $m$ is specified by three
physical parameters ($\tau_m$, $V_m$, $\lambda_m(t)$): 
$\tau_m$ represents the time instant at which the content enters the
system (i.e., it becomes available to the users); $V_m$ denotes the
average number of requests generated by the content; $\lambda_m(t)$ is
the ``popularity profile'', describing how the request rate for
content $m$ evolves over time.  In general, function $\lambda_m(t)$ satisfies
the following conditions: (positiveness)
$\lambda_m(t) \ge 0$, $\forall t$; (causality) $\lambda_m(t)=0$,
$\forall t<0$; 
(normalization) $\int_0^\infty \lambda_m(t) \diff t =1$.  We define the
{\em average life-span} of content $m$, $L_m$, as follows:
\mmm\mmm
\begin{equation}\label{eq:l}
L_m=\dfrac{1}{\int_{0}^\infty \lambda_m^2(t) dt}
\end{equation}\mmm
It will become clear later, while analyzing the performance of LRU  
in Sec.~\ref{sec:modelling}, why it is convenient to define 
the life-span of a content using the formula above. 
For now, to get an initial understanding of the definition, consider a
content with a uniform popularity profile $\lambda_m(t)=1/\delta$ for
$t\in[0,\delta]$. By computing~\eqref{eq:l}, we obtain $L_m=\delta$,
which is the intuitive value of life-span of such content.

Given the above parameters, our model assumes that the request process
for content $m$ is described by a time-inhomogeneous Poisson process
whose instantaneous rate at time $t$ is given by 
\(
V_m \cdot \lambda_m(t- \tau_m) 
\).

For the sake of simplicity, we assume that new contents become
available in the system according to a homogeneous Poisson process of
rate $\gamma$, i.e., time instants $\{\tau_m\}_m$ form a standard
Poisson process.  We refer to this model as Shot Noise Model (SNM),
since the overall process of requests arrival is known as a Poisson
shot-noise process~\cite{shot}. \autoref{fig:non-stationary}
illustrates an example of the request pattern generated by the
superposition of two ``shots'' corresponding to two contents having
quite different parameters.
We emphasize that the above Poisson assumptions, on the (instantaneous)
generation process of requests for each content,
and on the arrival process of new contents, are essentially introduced for the
sake of analytical tractability.  However, they are very well
justified by the experience gained from our traces, which show that it
is not really important to capture complex arrival dynamics at short
time-scales (see discussion about the results in
\autoref{fig:phit_vs_cachesize_shuffle_flip_trace_01}).
\tgifeps{6.5}{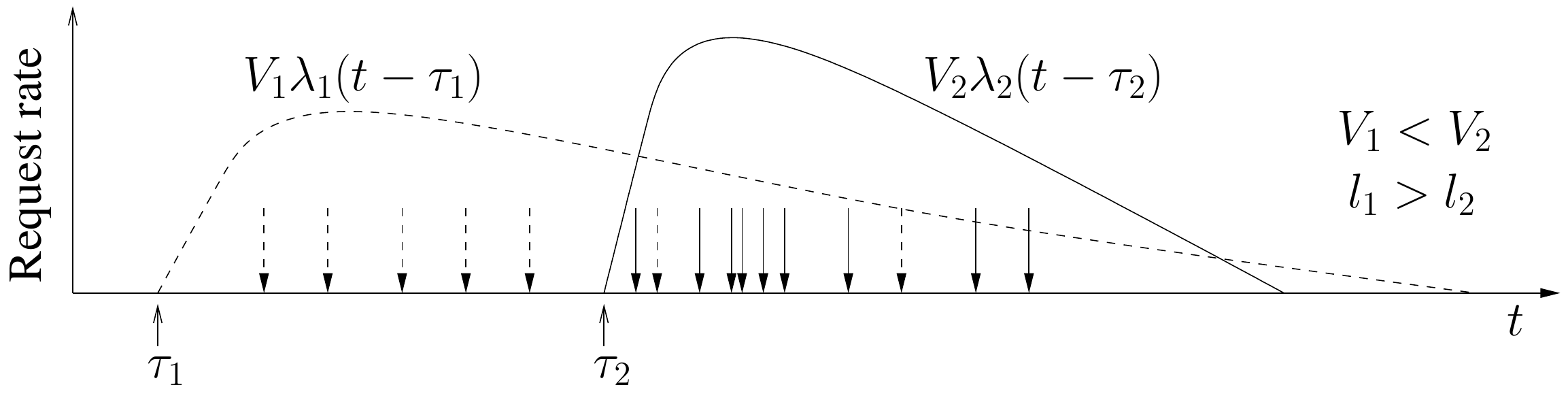}{Example of requests (denoted by arrows)
  generated by two contents with different catalogue insertion time
  ($\tau_1, \tau_2$), average number of requests ($V_1,V_2$) and profiles
  ($\lambda_1(t)$,$\lambda_2(t)$). \vspace*{-0.8cm}}

For a given content, the SNM requires us to specify its entire
popularity profile in the form of the function $\lambda_m(t)$, which,
given the difficulty in estimating popularity profiles from a trace,
could be considered as a limitation.  However, we have found that {\em
  it is not necessary to precisely identify the shape of
  $\lambda_m(t)$}. In fact, a simple first-order approximation,
according to which we just specify the average content life-span
$L_m$,
is enough to obtain accurate predictions of cache performance. In
other words, we can arbitrarily choose any reasonable function
$\lambda_m(t)$ with an assigned life-span $L_m$, and obtain almost the
same results in terms of cache performance (see the later discussion on \autoref{fig:phit_vs_cachesize_fertility_flip_trace_04}).
Finally, content heterogeneity is taken into account by associating
to every content, its life-span $L_m$, jointly with the (typically
correlated) average number of requests $V_m$.  This means that, upon
arrival of each new content $m$, we randomly choose (independently for
each content) the pair of parameters ($V_m$, $L_m$) from a given
assigned joint distribution.

\vspace*{-0.2cm}
\subsection{Validation of basic SNM}\label{subsec:valid_single}
To show how our traffic model can accurately capture
the temporal locality observed in real data, and 
its impact on cache performance, we introduce
a simple procedure to fit its parameters from a trace. 
 
For each given content in
the trace first we compute the total number of observed requests
$\hat{V}_m$, and then  we estimate the content life-span $\hat{L}_m$.
The interested reader can found  the details about how to estimate $\hat{L}_m$ from the traces in the preliminary version of this paper~\cite{ours_ccr}.
\begin{table}[t!]
\centering
 \tiny
\begin{tabular}{|c|c|c|c|c|c|c|}
\hline
Class & Classification rule   & Trace   & \%Reqs& \%Videos  & $\mathbb{E}[\hat{L}_m]$   & $\mathbb{E}[\hat{V}_m] $\cr\hline
\multirow{4}{*}{Class 0}  & \multirow{4}{*}{$\hat{V}_m<10$}       &\TONE\   & 74.60 & 98.588    &   -     & 1.41    \cr
        &                                                         &\TTWO\   & 72.64 & 98.401    &   -     & 1.42    \cr
        &                                                         &\TTHREE\ & 72.53 & 98.210    &   -     & 1.44    \cr
        &                                                         &\TFOUR\  & 67.30 & 97.778    &   -     & 1.49    \cr\hline
\multirow{4}{*}{Class 1}  & \multirow{3}{*}{$\hat{V}_m\geq 10$}   &\TONE\   & 2.34  & 0.044   & 1.14    & 86.4    \cr
        &                                                         &\TTWO\   & 2.71  & 0.083   & 1.09    & 76.2    \cr
        & \multirow{2}{*}{$\hat{L}_m \le 2$}                      &\TTHREE\ & 2.60  & 0.067   & 1.04    & 76.0    \cr
        &                                                         &\TFOUR\  & 2.81  & 0.077   & 1.06    & 74.0    \cr\hline
\multirow{4}{*}{Class 2}  & \multirow{3}{*}{$\hat{V}_m\geq 10$}   &\TONE\   & 1.72  & 0.069   & 3.36    & 41.9    \cr
        &                                                         &\TTWO\   & 3.43  & 0.125   & 3.34    & 50.7    \cr
        & \multirow{2}{*}{$2<\hat{L}_m\le 5$}                     &\TTHREE\ & 1.77  & 0.082   & 3.32    & 43.3    \cr
        &                                                         &\TFOUR\  & 2.01  & 0.093   & 3.41    & 48.0    \cr\hline
\multirow{4}{*}{Class 3}  & \multirow{3}{*}{$\hat{V}_m\geq 10$}   &\TONE\   & 1.49  & 0.041   & 6.40    & 59.5    \cr
        &                                                         &\TTWO\   & 1.84  & 0.070   & 6.31    & 44.9    \cr
        & \multirow{2}{*}{$5<\hat{L}_m\le 8$}                     &\TTHREE\ & 1.66  & 0.052   & 6.42    & 63.3    \cr
        &                                                         &\TFOUR\  & 1.64  & 0.062   & 6.45    & 60.3    \cr\hline
\multirow{4}{*}{Class 4}  & \multirow{3}{*}{$\hat{V}_m\geq 10$}   &\TONE\   & 1.39  & 0.062   & 10.53   & 36.9    \cr
        &                                                         &\TTWO\   & 2.96  & 0.128   & 10.86   & 39.6    \cr
        & \multirow{2}{*}{$8<\hat{L}_m\le 13$}                    &\TTHREE\ & 1.33  & 0.066   & 10.62   & 39.5    \cr
        &                                                         &\TFOUR\  & 1.75  & 0.103   & 10.65   & 37.8    \cr\hline
\multirow{4}{*}{Class 5}  & \multirow{3}{*}{$\hat{V}_m\geq 10$}   &\TONE\   & 18.46 & 1.196   & 24.61   & 25.7    \cr
        &                                                         &\TTWO\   & 16.41 & 1.193   & 19.29   & 25.3    \cr
        & \multirow{2}{*}{$\hat{L}_m > 13$}                       &\TTHREE\ & 20.11 & 1.523   & 28.19   & 25.8    \cr
        &                                                         &\TFOUR\  & 24.49 & 1.887   & 24.59   & 28.1    \cr\hline
\end{tabular}
\caption{Model parameters for each content class. $\hat{L}_m$ is evaluated in days. The numbers of requests and videos for classes 0--5 have been normalized separately for each trace.}\label{tab:desc-dataset}
\vspace*{-0.2cm}
\end{table}
Contents are then partitioned into $6$ classes ($0,\ldots,5$), on the basis of
previous values $\hat{V}_m$ and $\hat{L}_m$. 
Class~0 comprises all the contents with small number of requests ($\hat{V}_m < 10$), 
for which  we cannot derive a
reliable estimate of their life-span. 
Contents in classes $1$ to $5$ contain contents with $\hat{V}_m \geq 10$ which, 
as reported in \autoref{tab:desc-dataset}, are partitioned
according to $\hat{L}_m$ (measured in days). For each class,
\autoref{tab:desc-dataset} reports, for four different traces, 
the percentage of total requests
attracted by the class, the percentage of videos belonging to it,
and the average values $E[\hat{L}_m]$ and $E[\hat{V}_m]$.
for the videos in the class.

From \autoref{tab:desc-dataset}, we observe that: (i) The values
related to each class are quite similar (with differences of at most $20\%$)
across the considered traces. This is significant, because it suggests that
our broad classification captures some invariant properties of the
considered VoD traffic. (ii) Contents in Class~1 (having $\hat{L}_m < 2$~days)
represent roughly $0.07\%$ of the total number of contents ($4\%$ of contents in Classes 1-5),
but account for approximately $2.5\%$ of all requests ($10\%$ of requests 
originated by  contents in Classes 1-5). 
These contents exhibit the larger degree of temporal locality, and we will see
that their impact on cache performance is crucial, despite
the fact that they represent a rather small fraction of the traffic.  
(iii) Contents in Class~5 have a life-span comparable with the
trace length, therefore their measured value $\hat{L}_m$ is expected to be
strongly affected (i.e., underestimated) by border effects due to the
finiteness of the trace.
We chose not to attempt to characterize the temporal locality of contents belonging to either
Class~0 (too few requests) and Class~5 (unreliable estimate of
life-span), by using the SNM. We therefore treat these contents as
if their popularity was stationary (like in the IRM), and generate
their requests uniformly in the considered time horizon.
We emphasize that, with this choice, we miss the opportunity
to capture the temporal locality of a large fraction of contents.
On the other hand, by so doing we obtain a conservative
prediction of cache performance.
 
As such, only the requests for contents falling in Classes $1$ to $4$ 
are generated according to the SNM model,
assuming a common ``shape'' $\lambda(t)$ for the
popularity profile, chosen from the profiles 
listed in~\autoref{tab:profiles2}.  For each content class, the 
shape parameter $\delta$ is chosen to match the
average life-span $\mathbb{E}[\hat{L}_m]$.  Moreover, for each class
modeled by the SNM, the distribution of request volumes was matched to the
corresponding empirical distribution observed in the trace.

\begin{table}[!tb]
  \tiny
  \centering
  \begin{tabular}{|l|c||c|}
    \hline
    Profile & $\lambda(t)$ &  $L$ \\
    \hline
       Uniform & ${1}/{\delta}$ for $t\in[0,\delta]$ & ${\delta}$\\
    Exponential & $({1}/{\delta}) e^{-t/\delta}$ for $t\geq 0$ & $2\delta$\\
    Power law ($\zeta>1$) &
    $\dfrac{\zeta-1}{\delta}\left(\dfrac{t}{\delta}+1\right)^{-\zeta}$ for $t\geq 0$&
    $\dfrac{\delta(2\zeta-1)}{(\zeta-1)^2}$\\
    \hline
  \end{tabular}
  \caption{Popularity profile and corresponding life-span $L$.}
  \label{tab:profiles2}
  \vspace*{-0.4cm}
\end{table}

\tgifeps{5.5}{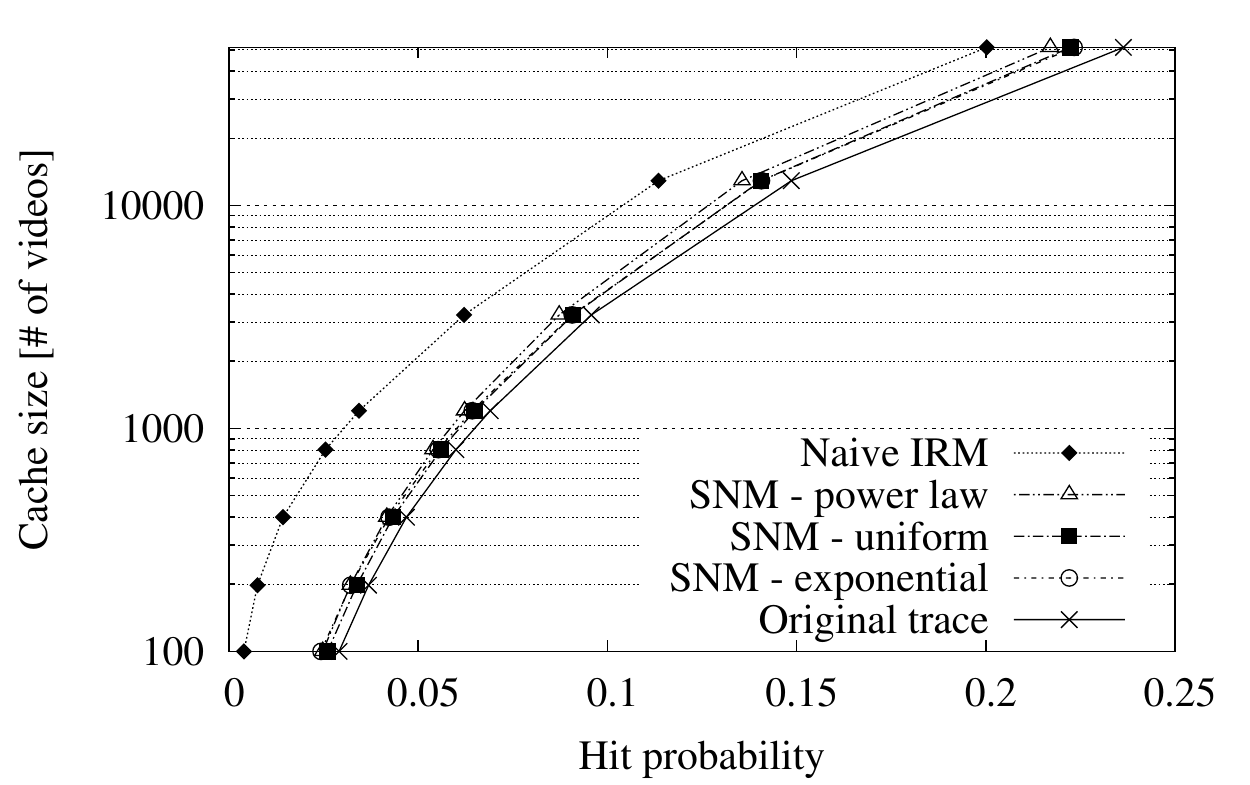}{Cache size vs hit probability
under LRU, for \TFOUR. \vspace*{-0.5cm}}

We generated a
synthetic request trace using the parameters estimated as above,
and fed it
 to a cache implementing the LRU policy.
Fig.~\ref{fig:phit_vs_cachesize_fertility_flip_trace_04} reports the
cache size required to achieve a desired hit probability, using \TFOUR
~(similar results were obtained with the other traces).  For
comparison, we report also the results obtained with the original trace
and those obtained by its completely shuffled version representing the ``naive IRM'' approach.
We observe that the results obtained by applying the fitted SNM (using
either uniform,  exponential or power-law shape with
$\zeta=3$) are very close to those obtained with the original
unmodified trace, and that the shape chosen for the popularity profile
has little impact on the results.

In summary, Fig.~\ref{fig:phit_vs_cachesize_fertility_flip_trace_04} 
shows that our SNM provides an accurate prediction of cache performance, despite the heavy
simplifications adopted in the parameters' identification. 
We expect that even more accurate predictions
could be achieved by improving the fitting procedure, or by using much
longer traces (if available).

\vspace*{-0.3cm}

\subsection{Extension of SNM to cache networks} \label{subsec:snm_multi}

We now extend the basic SNM introduced in
Sec.~\ref{subsec:shotonecache} to handle the case of multiple
interconnected caches, in which edge caches receive the requests
generated by subsets of users 
possibly having quite different interests from 
one ingress point to another (geographical locality).
The extension of the basic model can be, in principle, carried out in
its most general form by associating to every content $m$ and ingress
point $i\in I$, a tuple $(\tau_{m,i}$, $V_{m,i}$,
$\lambda_{m,i})$, so that, at ingress point $i$, requests for content
$m$ arrive according to an inhomogeneous Poisson process, whose
instantaneous rate at time $t$ is given by
\(
 V_{m,i} \cdot \lambda_{m,i}(t-\tau_{m,i})
\).
While this approach is fairly general,  it is not very
practical, because the total number of
parameters to be specified may become very large.
An alternative is to make the following simplifications.
First, we assume that the instants at which content $m$ starts to be
available in the system at the various ingress points ($\tau_{m,i}$)
are equal, i.e., $\tau_{m,i} = \tau_{m}$. This is well justified since
in most systems contents are enabled to be globally available to all
users at the same time. Second, the popularity of a given content $m$
is assumed to follow the same profile $\lambda(t)$ at every ingress point. 
Together with the previous consideration regarding the negligible impact of
the particular shape of the popularity profile, 
we represent each content $m$ by: i) its (global) starting
point of availability $\tau_m$; ii) its (global) life-span $L_m$; iii) a
set $\{V_{m,i}\}_i$ of (local) parameters denoting the volumes of
requests attracted by content $m$ at different ingress points.

Our simplifications are realistic for cache networks covering
limited geographical areas.  We remark that, when considering content
distribution systems spanning large areas, including very different
time zones (i.e.\ at world-wide scale), temporal profiles (especially
for contents having small life-span) should not be considered synchronized,
but properly modified 
to take jointly into account geographical locality and diurnal patterns. 

Finally, to specify the volumes of requests generated by contents at the
various ingress points ($V_{m,i}$), we adopt the following approach: for each content $m$, we
assign a global volume $V_m$, denoting the total number of requests 
generated in the whole system. Then, we specify $V_{m,i}$ as
\(
V_{m,i}= V_m \cdot  p_{i,m}
\),
where $p_{i,m}$ represents the fraction of requests for content $m$
arriving at ingress point $i$. By construction $p_{i,m} \geq 0$ and
$\sum_i p_{i,m} = 1$.
This approach allows a large degree of flexibility in
describing the geographical locality of contents. For example, we can
obtain the case in which geographical locality is negligible, by setting
$p_{i,m} = p_i$, for every $m$; here $p_i$ represents the ``relative
mass'' of requests arriving at ingress point $i$ (i.e., the
fraction of all requests generated by the corresponding set of users).
At the other extreme, we can represent the case of complete geographical
locality by setting $p_{i,m} \in \{0,1\}$ (i.e., setting each content
to be requested at only one particular ingress point).

\mmm\mmm\mmm
\subsection{Validation of SNM for cache networks} \label{subsec:validate_multi}
To assess the validity of our simplifying assumptions, according to which
the popularity evolution of each content is ``synchronized'' across different
ingress points (i.e., $\tau_{m,i} = \tau_{m}$ and $\lambda_{m,i} =
\lambda_{m}$, $\forall i$), we evaluated the degree of temporal overlap
between the sequence of requests received by a content in different traces
of our data set. 
In particular, given a pair of traces, we computed the following metric
for each content $m$ which appears in both traces:
\mmm\mmm
\[
\textrm{time-overlap-fraction}(m) = \frac{\textrm{$|$intersection of
    life-spans$|$}}{\textrm{$|$union of life-spans$|$}}\mmm
\]
Note that, by definition, the above metric takes 
values in $[0,1]$, where $0$ represents no overlap and $1$ is produced by
identical and perfectly overlapped life-spans.  
For example, the case of two equal life-spans, shifted in time by 20\%, leads to
$\textrm{time-overlap-fraction} = 0.8/1.2=0.67$.

\begin{figure}[t!]
  \begin{subfigure}[t]{0.48\columnwidth}
    \hspace*{-0.6cm}
    \includegraphics[width=1.15\linewidth]{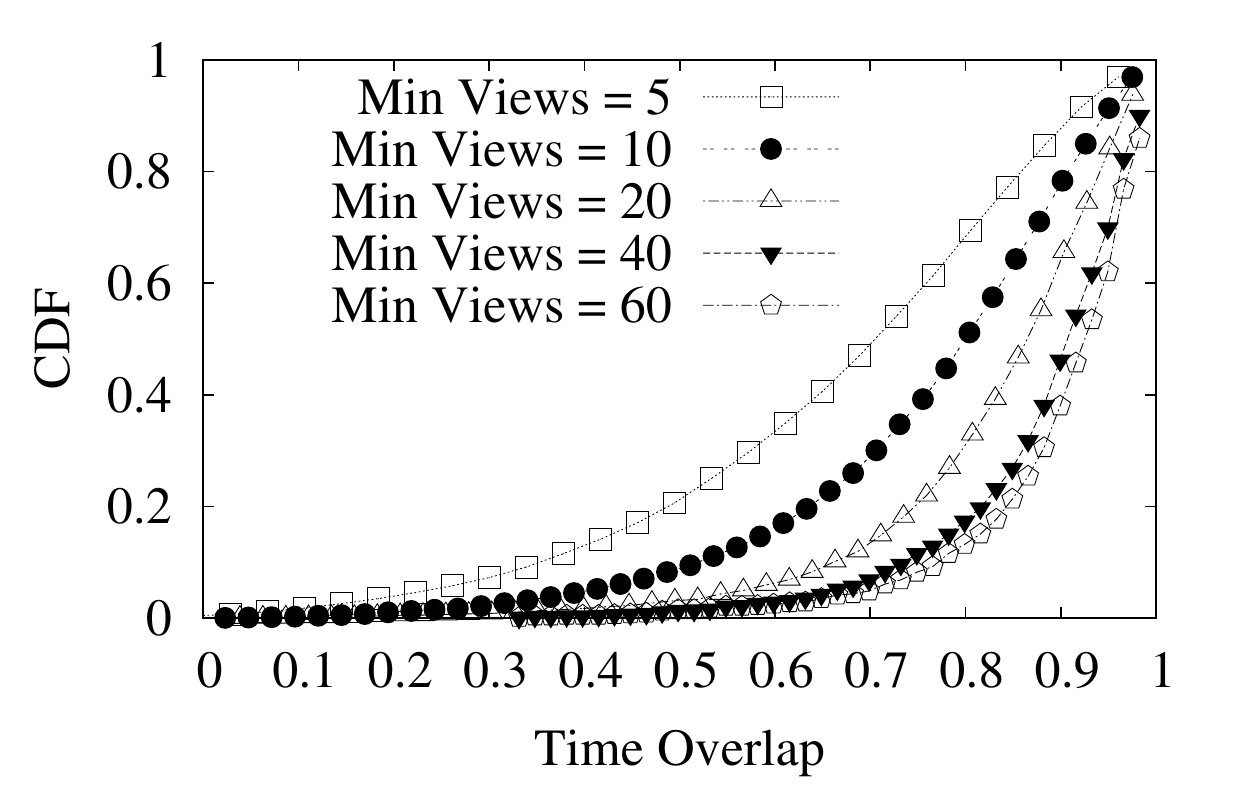}
    \caption{\TSIX\ and \TSEVEN\ (same country).}
    \label{fig:overlap_cdf_FWPDF-FWPUL}
  \end{subfigure}
  \hspace*{0.1cm}
  \begin{subfigure}[t]{0.48\columnwidth}
    \hspace*{-0.4cm}
    \includegraphics[width=1.15\textwidth]{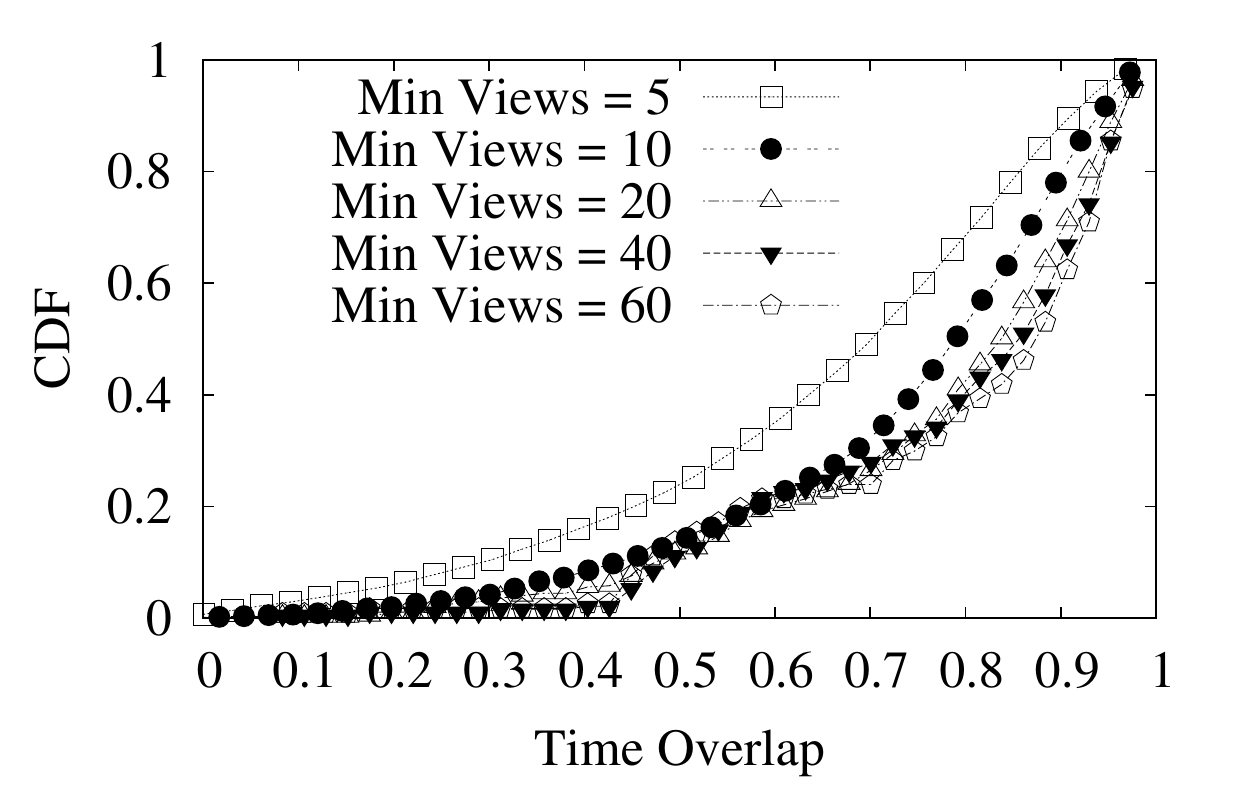}
    \caption{\TTHREE\ and \TEIGHT\ (different countries).}
    \label{fig:overlap_cdf_FWPUL-POLI}
  \end{subfigure}
  \caption{Time overlap for different pairs of traces.}
  \vspace*{-0.3cm}
\end{figure}

Fig.s~\ref{fig:overlap_cdf_FWPDF-FWPUL}
and~\ref{fig:overlap_cdf_FWPUL-POLI}, respectively report the CDFs of
the time-overlap-fraction for \TSIX\ and \TSEVEN\ (collected from residential
networks in the same country) and for \TTHREE\ and \TEIGHT\ (collected from
residential networks in different countries). To get additional insight,
we obtained a separate CDF for the contents
which receive at least a certain minimum number of requests 
in both traces (this is the parameter ``Min Views'' reported
in the figures, ranging from 5 to 60). 
We observe that the degree of synchronization increases
with the content popularity.
This can be in part due to the fact 
that the  life-span interval of a content resulting from a trace is affected randomly by border 
effects, which smooth out as the request volume increases. Indeed,
even when the popularity evolution of a  content  is perfectly synchronized across different ingress points, i.e.\  
$\tau_{m,i}=\tau_{m}$ and $\lambda_{m,i} =\lambda_{m}$, the overlap observed in the actual sequence of 
requests can be small for contents with few requests.
Nevertheless, we observe quite a strong degree of synchronizing
in both Fig.s~\ref{fig:overlap_cdf_FWPDF-FWPUL} and~\ref{fig:overlap_cdf_FWPUL-POLI}.
For example, only about $25\%$ of the contents receiving more than 
just $10$ requests show a time-overlap-fraction smaller 
than $0.67$.
\begin{table}[!b]
\centering
\tiny
\begin{tabular}{|c|c|c|c|c|c|c|}
\hline
\multicolumn{2}{|c|}{Final class}             & Class 1   & Class 2 & Class 3 & Class 4 & Class 5 \\
\hline\hline
\multicolumn{2}{|c|}{Initial class} &\multicolumn{5}{c|}{ \TSEVEN }\\

      \hline
 &   Class 1         & {\em 0.69}   & 0.23  &  0.04 &  0.02 &    0.02 \cr
 & Class 2 & 0.11 & {\em 0.64} & 0.16 & 0.05 & 0.04 \\
 \TSIX                                  
 &   Class 3         &  0.03 & 0.21 & {\em 0.44} & 0.16 & 0.16 \\
 &   Class 4         &   0.01 & 0.07 & 0.18 & {\em 0.35} & 0.39 \\
 &   Class 5         &   0.00   & 0.00 & 0.01 & 0.02 & {\em 0.97}\cr
 \hline
\hline
&&\multicolumn{5}{c|}{ \TEIGHT }\\
\hline
  & Class 1 & {\em 0.86} & 0.06 & 0.04 & 0.00 & 0.01\\
  &   Class 2         &0.30 & {\em 0.56} & 0.08 & 0.03 & 0.03 \cr
\TTHREE
  & Class 3 & 0.10 & 0.27 & {\em 0.39} & 0.13 & 0.11 \\
   &   Class 4         & 0.02 & 0.12 & 0.26 & {\em 0.44} & 0.16 \\
   & Class 5 & 0.00 & 0.01 & 0.02 & 0.03 & {\em 0.94} \\
\hline
\end{tabular}
\caption{{Cross-classification of contents for two
    different pairs of traces.
    }} \label{tab:video-perc}
\vspace{-0.3cm}
\end{table}

To show how the life-span of a content measured in different
traces can vary in size, \autoref{tab:video-perc} reports,
for the same two pairs of traces considered before, 
the fraction of contents that, given an initial classification
in one trace (the row), were classified into a given class (the column)
in the second trace (we refer to the definition of classes 
in Table~\ref{tab:desc-dataset}). By construction, each row
in the table sums to one.
We observe that the majority of contents are classified
into the same class in both traces. Moreover,
the fraction of contents for which the class index
differs by more than one is negligible. {Note that contents
whose class index differs exactly by one (i.e., the cells in the 
first diagonal above or below the main diagonal) 
should be taken with care, because their life-spans 
might be close to the threshold value separating
two neighboring classes, which can easily lead to a  
misclassification.}

Having observed that the popularity evolution of contents 
appears to be well synchronized across the traces (both in terms of 
overlap and size of the corresponding life-span intervals),
we performed one more experiment to validate our modeling approach, 
to check how the non-perfect synchronization existing in real 
traces can affect the resulting cache performance.
\begin{figure}[tb!]
    \centering \includegraphics[width=5.5cm,
      clip=true]{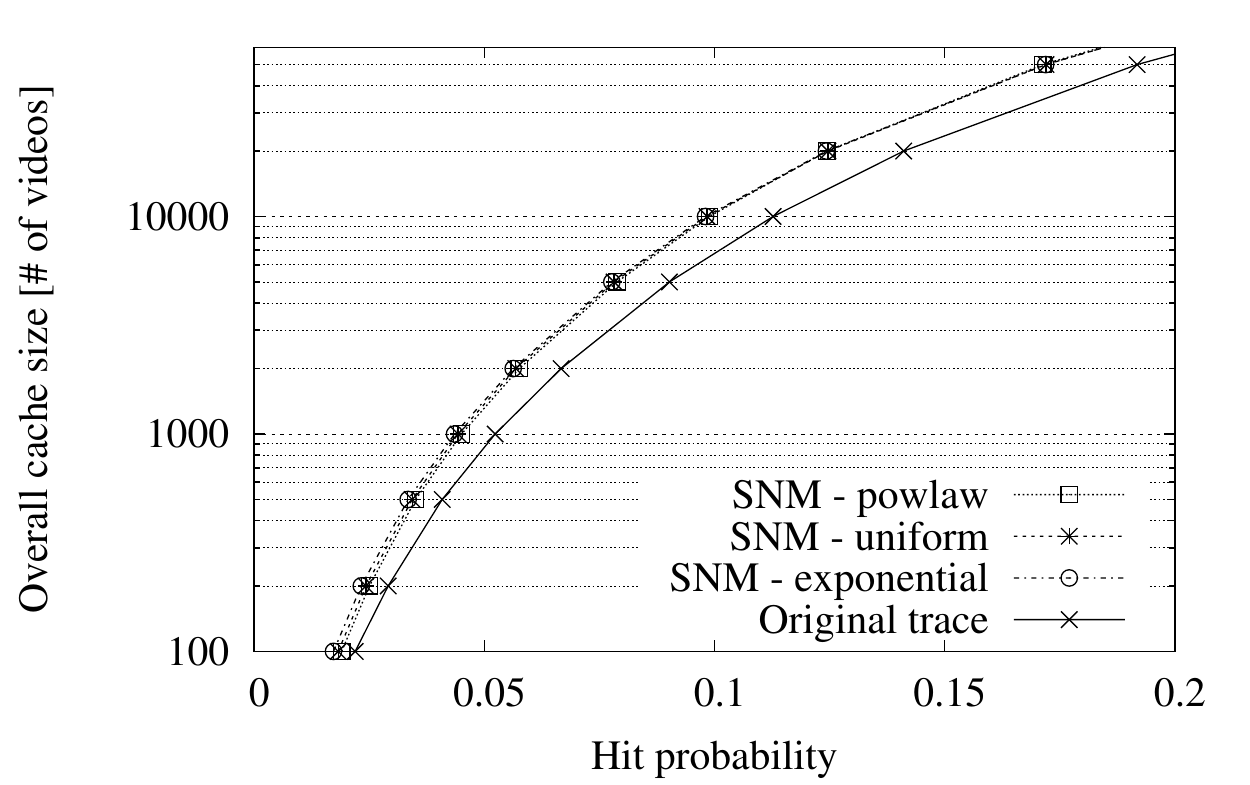}
    \vspace{-0.1cm}
    \caption{Overall cache size ($60\%$ in the root, $20\%$ in each
      leaf) vs hit probability, in a cache network fed either by  
    real traces or by synchronized synthetic traffic.}
    \label{fig:p-pul-pdf}
    \vspace*{-0.5cm}
\end{figure}
In particular, we considered a simple cache network composed by one 
root and two leaves. We assume that the root takes $60\%$ of
the overall cache size and each leaf takes $20\%$ of it.
Then we fed the left leaf by \TONE and the right leaf by \TTHREE.
These traces where chosen because they contain similar request volumes and 
have a large temporal intersection.
In the case of a miss, requests are forwarded to the root. 
Contents are replicated on all caches traversed by a request.

To obtain an easily tractable traffic model for this scenario,
relying on the synchronization assumption, we proceeded as 
follows: we merged the two traces, and derived a unique 
SNM from the combined trace, using the same fitting procedure described in 
Sec.~\ref{subsec:valid_single}. The requests in the synthetic trace produced by the SNM 
are then randomly distributed between the two leaves, in proportion to the 
number of requests in the original traces. Note that, by so doing, 
contents are assumed to be perfectly synchronized 
between the two ingress points.

Fig.~\ref{fig:p-pul-pdf} shows the overall cache size needed to obtain 
a given hit probability. The curve labeled ``Original trace'' refers
to the case in which the network is fed by real traffic traces,
whereas the other curves refer to synthetic traffic traces
produced by the fitted SNM, using different shapes for 
the popularity profile. We observe that, despite all
approximations and simplifying assumptions, 
our traffic model provides good predictions of cache performance, 
being the required cache size overestimated by a factor always lower 
than 2.

\section{Analysis of LRU under SNM}\label{sec:modelling}

The Shot Noise Model introduced and validated in Sec.~\ref{sec:analysis}
is simple enough to permit developing accurate
analytical models of classic caching policies. 
In particular, in this section we show how the  
Least Recently Used (LRU) caching policy can be analyzed under the traffic produced by the 
SNM, by extending a technique known as Che's
approximation~\cite{che02}. 
Note that, to ease the readability, the extension of our 
analysis to cache networks was moved to Appendix~\ref{sec:nets}.

\subsection{The single cache case}\label{subsec:modelsingle}
Consider a cache capable of storing $C$ distinct contents. Let
$T_C(m)$ be the time needed for $C$ distinct contents, not including
$m$, to be requested by users. $T_C(m)$ therefore gives the {\em cache
  eviction time} for content $m$, i.e., the time since the last
request for content $m$, after which content $m$ is evicted from the
cache.  Of course $T_C(m)$ is a stochastic variable whose distribution
typically depends on the considered content $m$; Che's approximation
is based on the simplifying assumption that the cache eviction time
($T_C(m)$) is {\em deterministic} and {\em independent} of the considered content
($m$).  This assumption has recently been given a theoretical
justification in~\cite{Roberts-ITC}, where it was shown that, under
IRM with a Zipf-like (static) popularity distribution, the coefficient
of variation of $T_C(m)$ tends to vanish as the cache size grows.
Furthermore, the dependence of the eviction time on $m$ becomes
negligible when the content catalogue is sufficiently large. Moreover,
in~\cite{Roberts-ITC} authors discover that Che's approximation is
also surprisingly accurate in critical cases (small catalogue, very
skewed popularity law). The arguments used in~\cite{Roberts-ITC} are
easily extended to our non-stationary traffic model when the product
$\gamma \cdot \mathbb{E}[L_m]$ (the average number of concurrently
``active'' contents) and $C$ are sufficiently large.

We start our analysis by considering the single-class case, in which
the popularity profile ($\lambda(t)$) is the same for all
contents, being characterized by the average content life-span $L$.
The request volumes ($V_m$) are assumed to be i.i.d.\ random variables, distributed as $V$, with
$\mathbb E[V]<\infty$.  Finally, we define
$\phi_V(x)=\mathbb{E}[e^{xV}]$ to be the moment generating function of
$V$ and $\phi_V'(x)=\mathbb{E}[V e^{xV}]$ its first derivative, such
that $\phi_V'(x)\geq 0$ for any $x$.
Heterogeneity of the contents' popularity is handled by a
multi-class extension that will be described later in Sec.~\ref{sec:ext}.

\subsubsection{Main result for the single-class model}\label{sec:main}
In the case of a single class of contents, the application of Che's
approximation to analyze LRU performance under the SNM leads to the
following fundamental result:

\begin{teorema}\label{theo:1}
Consider a cache of size $C$ implementing LRU policy, operating under
a SNM request arrival process with total stochastic intensity:\mmm\mmm
\[
  \Lambda(t)=\sum_{m:\tau_m<t} V_m \lambda(t-\tau_m)\mmm
\]
representing $\tau_m$ points of an homogeneous Poisson process with
intensity $\gamma$ and $V_m$ i.i.d. random variables.  Under the Che's
approximation, the hit probability is given by:\mmm
\begin{equation}\label{eq:phit}
  p_{\text{hit}} = 1 - \int_0^{\infty}\lambda( \tau) \
  \frac{\phi_V' \left(-\int_ {0}^{T_C} \
      \lambda(\tau - \theta)d \theta \right)}  { \mathbb{E}[V] }  d \tau \mmm
\end{equation}
where $T_C$ is the only solution to equation:
\begin{equation}\label{eq:c2}
  C=\gamma \int_{0}^\infty 1 - \phi_V \
  \left(-\int_ {0}^{T_C} \lambda( \tau - \theta) d\theta \right) d \tau
\end{equation}
\end{teorema}
\begin{IEEEproof}
The proof is reported in Appendix \ref{app:theo1}.
\end{IEEEproof}

\subsubsection{Small-cache regime}\label{sec:sc}
For small cache sizes, it is possible to derive a closed-form
approximation of \equaref{phit} and \equaref{c2} as follows:
\begin{corollario}\label{coro:psmall}
If ~$\mathbb E[V^2]<\infty$, for small cache sizes the hit probability
is approximated by:\mmm\mmm
\begin{equation}\label{eq:psmall}
  p_{\text{hit}}\approx\dfrac{T_C}{L} \frac{\mathbb E[V^2] }{\mathbb E[V]}\mmm
\end{equation}
where $T_C$ derives from equation:\mmm\mmm
\begin{equation}\label{eq:cts}
C=\gamma \mathbb E[V] T_C
\end{equation}
\end{corollario}
\mmm\begin{IEEEproof}
The expression in~\eqref{eq:psmall} is obtained from~(\ref{eq:phit})
by approximating $\int_0^{T_C} \lambda(\tau-\theta) d\theta $ with
$\lambda(\tau)T_C \ll 1$, and by locally approximating $\phi_V'(x)$,
in a right neighborhood of $x=0$, with its linear Taylor
expansion: $\phi_V'(x)=\mathbb{E}[Ve^{-xV}] \approx \mathbb{E}[V] -x
\mathbb{E}[V^2]$. At the same time, \eqref{eq:cts} can be obtained by
linearizing  \eqref{eq:c2}  for small $T_C$.
\end{IEEEproof}

By combining~\eqref{eq:psmall} and~\eqref{eq:cts}, we obtain the 
important result:
\begin{corollario} The hit probability under small-cache regime can be approximated as\mmm\mmm\mmm\mmm
\begin{equation}\label{eq:psmall2}
  p_{\text{hit}}\approx \dfrac{C}{\gamma L} \frac{\mathbb E[V^2]} {
    \mathbb {E}^2[V]}
\end{equation}
\end{corollario}

\begin{remark}\label{remark:2}
From~\eqref{eq:psmall2}, we gain the fundamental insight that, {\em
  when the cache size is small, relatively to the catalogue size, the
  hit probability of LRU under SNM traffic is insensitive to the
  detailed shape of the popularity profile, being inversely proportional to  
  the average content life-span $L$}.  Moreover, the dependency of
cache performance on the requests volume distribution ($V$) is
mediated by only the first two moments of it, through the ratio $\mathbb
E[V^2]/{\mathbb E}^2[V]$. Finally, according to~\eqref{eq:psmall2}, the
hit probability increases linearly with the cache size.  We will
show in Sec.~\ref{sec:validation} that \equaref{psmall2} also provides
a satisfactory approximation for quite large cache sizes, suggesting
the practical relevance of this simple formula.
\end{remark}

\vspace*{-0.4cm}
\subsubsection{Large-cache regime}\label{sec:hit:large}
As the cache size ($C$) tends to infinity, a closed-form expression for the asymptotic hit
probability, denoted by $p_{\text{hit},  \infty}$, can be derived from (\ref{eq:phit}) 
by making $T_C\to \infty$.

\begin{corollario}\label{cor:lc}
  For large cache sizes,\mmm\mmm
  \begin{equation}\label{eq:pinf}
    p_{\text{hit}, \infty}=1 - \frac{1-\phi_V(-1)}{\mathbb{E}[V]}=
    1-\dfrac{1}{\mathbb E[V]}+ \dfrac{\mathbb E[e^{-V}]}{\mathbb E[V]}\mmm
  \end{equation}
\end{corollario}

Observe that the dependency on the content popularity profile
($\lambda(t)$) is completely washed out in~\eqref{eq:pinf}. Thus, the
impact of temporal locality on cache performance (in particular, the
popularity profile) tends to vanish as the cache size increases.  This
fact can be easily explained by observing that, for arbitrarily large
cache sizes, the first request for a content necessarily produces a
miss, whereas all subsequent requests lead to a hit. 

\begin{remark}\label{remark:3} 
As the cache size increases and~\eqref{eq:psmall2} degrades in its
accuracy, the hit probability tends to be affected by the detailed
shape of the temporal profile as well as the distribution of the requests
volume (as predicted by \eqref{eq:phit}). However, the overall impact
of temporal locality on the cache performance decreases, up to the
point of completely vanishing when the time scale of cache dynamics
(governed by the eviction time ($T_C$)) becomes larger than the content
life-span ($L$) (see \eqref{eq:pinf}).
\end{remark}

\vspace*{-0.3cm}
\subsubsection{Extension to multi-class scenario}~\label{sec:ext}
The single-class model in Sec.~\ref{subsec:modelsingle} can be
extended to consider the more realistic scenario in which contents are
partitioned into $K$ classes. We assume that each class is
characterized by a different popularity profile ($\lambda_k(t)$), with
an associated average content life-span $L_k$, and a request volume
$V_{(k)}$, for $1\le k\le K$.
Similarly to the single-class case, we assume $\mathbb
E[V_{(k)}]<\infty$ and we define $\phi_{V_{(k)}}(x)$ as the moment
generating function for $V_{(k)}$.
We can formalize the multi-class scenario by assuming that every
generated content $m$ is assigned a random mark $W_m$, representing
the class the content belongs to, taking values in $\{1, \ldots,
K\}$. Assuming $\{W_m \}_m$ to be i.i.d.~random variables, the total
stochastic intensity of the request process at time $t$ is given by:\mmm\mmm\mmm\mmm
\[
\Lambda(t)=\sum_{m:\tau_m<t} V_m \lambda_{W_m}(t-\tau_m)\mmm\mmm
\]
Under this assumption, we can state the following:
\begin{teorema}~\label{theo:mcLRU}
Consider a cache of size $C$, implementing the LRU policy, and
operating under a multi-class SNM model with total stochastic
intensity: \( \Lambda(t)=\sum_{m} V_m \lambda_{W_m}(t-\tau_m) \).
Extending Che's approximation, the hit probability is given by:
\vspace*{-3mm}
\begin{equation}\label{phitformulamulticlass}
\hspace{-3mm}
p_{\text{hit}}=1-\sum_{k=1}^K\!\Prob\{W_1=k\} \!\! \int_0^{\infty}
\!\!\! \lambda_k(\tau) \frac{\phi_{V_{(k)}}'\!\! \left(-\int_ {0}^{T_c}
    \lambda_k(\tau-\theta)d\theta \right)}{\mathbb{E}[V_{(k)}]} \diff \tau \vspace*{-2mm}
\end{equation}
where $T_C$ is the only solution to equation:
\vspace*{-1mm}
\begin{equation}\label{Cformulamulticlass}
\hspace{-5mm}
C \!=\! \gamma \! \int_{0}^{\infty}\!\!\Big[ 1- \sum_1^K
  \Prob\{W_1\!=\!k\}\cdot\vspace{-6mm} 
  {\phi_{V_{(k)}}\!\left(-\!\!\int_ {0}^{T_C}
     \!\!\! \lambda_k(\tau-\theta) d\theta \! \right)}\Big] \!\diff \tau
\end{equation}
\end{teorema}
\vspace{5mm}
The proof for Theorem~\ref{theo:mcLRU} (not reported here for the sake
of brevity) follows the same lines as in the proof of
Theorem~\ref{theo:1}.  Furthermore, when the cache size becomes small,
it is possible to derive a closed-form approximation
of \eqref{phitformulamulticlass} and \eqref{Cformulamulticlass}:

\begin{corollario}\label{coro:psmall-multi}
If $\mathbb E[V_{(k)}^2]<\infty$ for any $1\le k\le K$, for small cache sizes the hit
  probability can be approximated as:\mmm\mmm
\begin{equation}\label{eq:multipsmall}
  p_{\text{hit}}\approx T_C\sum_{k=1}^K\!\Prob\{W_1=k\}\dfrac{1}{L_k} \frac{\mathbb E[V_{(k)}^2] }{\mathbb E[V_{(k)}]}
\mmm
\end{equation}

where $T_C$ derives from equation:
\(
C=\gamma \mathbb E[V] T_C
\). 
\end{corollario}

\begin{remark}\label{remark:4}
When cache size is small, the hit probability in the multi-class case
is given by a weighted sum of contributions, related to the various classes, 
where each contribution is 
inversely  proportional to the average life-span of the corresponding class and proportional  to 
the ratio ${\mathbb E[V_{(k)}^2] }/{\mathbb E[V_{(k)}]}$.  

\end{remark}

\vspace*{-0.2cm}
\subsubsection{Diurnal patterns and cache invariance}~\label{sec:diurnal}
From our model we can derive an analytical explanation of why diurnal
variations in the aggregate arrival rate of requests, such as those
illustrated in Fig.~\ref{fig:reqs_vs_time_week_profile_trace_03}, have
no impact on the hit probability. 

Diurnal variations in the intensity of the arrival process of requests
can be obtained from the resulting effect of an envelope-modulation
applied to 
all its constituent components (i.e., the shots
associated to individual contents).  In fact, we can obtain any
desired modulation in the total intensity of the arrival process by
starting from a stationary sequence of content requests, and properly
diluting/densifying the associated timestamps over time, whilst
preserving the ordering of the requests.
\begin{figure*}[t!]
  \centering
  \begin{subfigure}[t]{0.32\textwidth}
    \centering
    \includegraphics[width=1.05\linewidth]{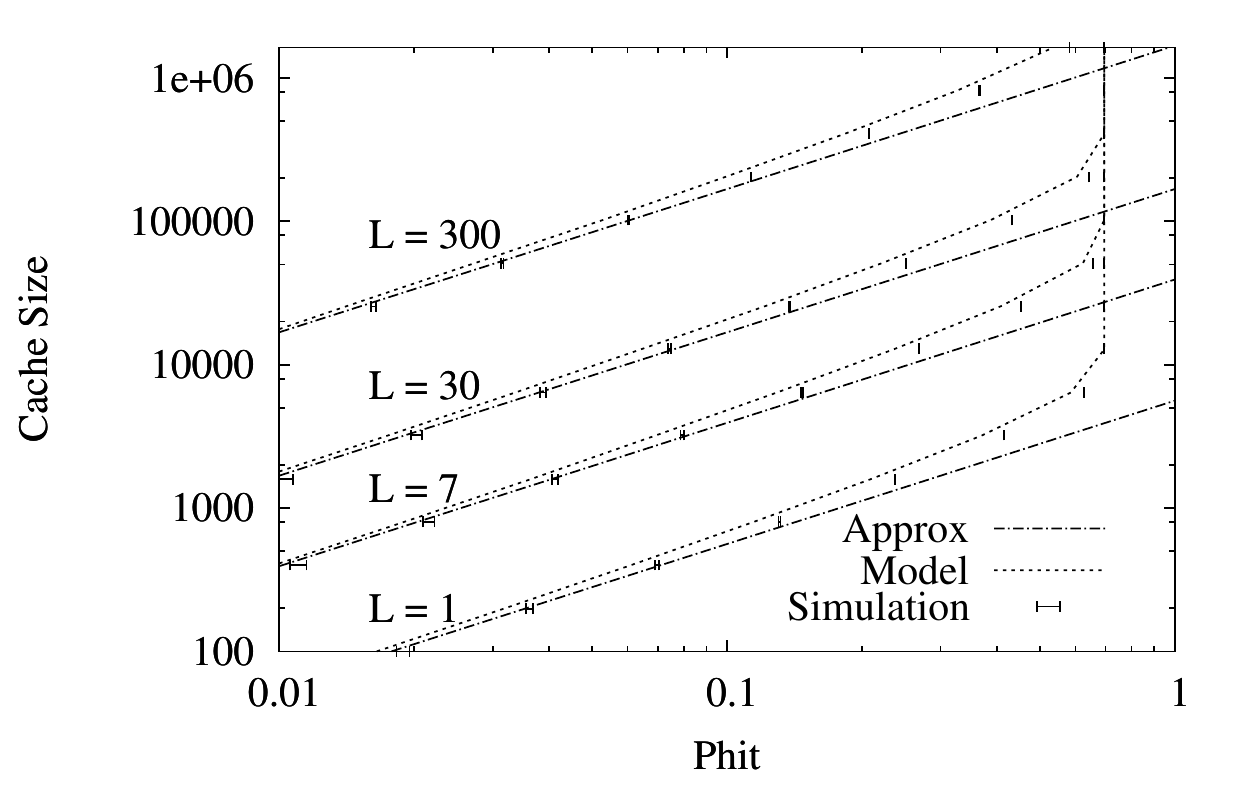}
    \caption{The impact of varying the average content life-span $L$
     (expressed in days).}
     \label{fig:cachesize_vs_phit_uniform_mod-sim_log_approx}
  \end{subfigure}%
  \hspace*{0.1cm}
  \begin{subfigure}[t]{0.32\textwidth}
    \centering
    \includegraphics[width=1.05\linewidth]{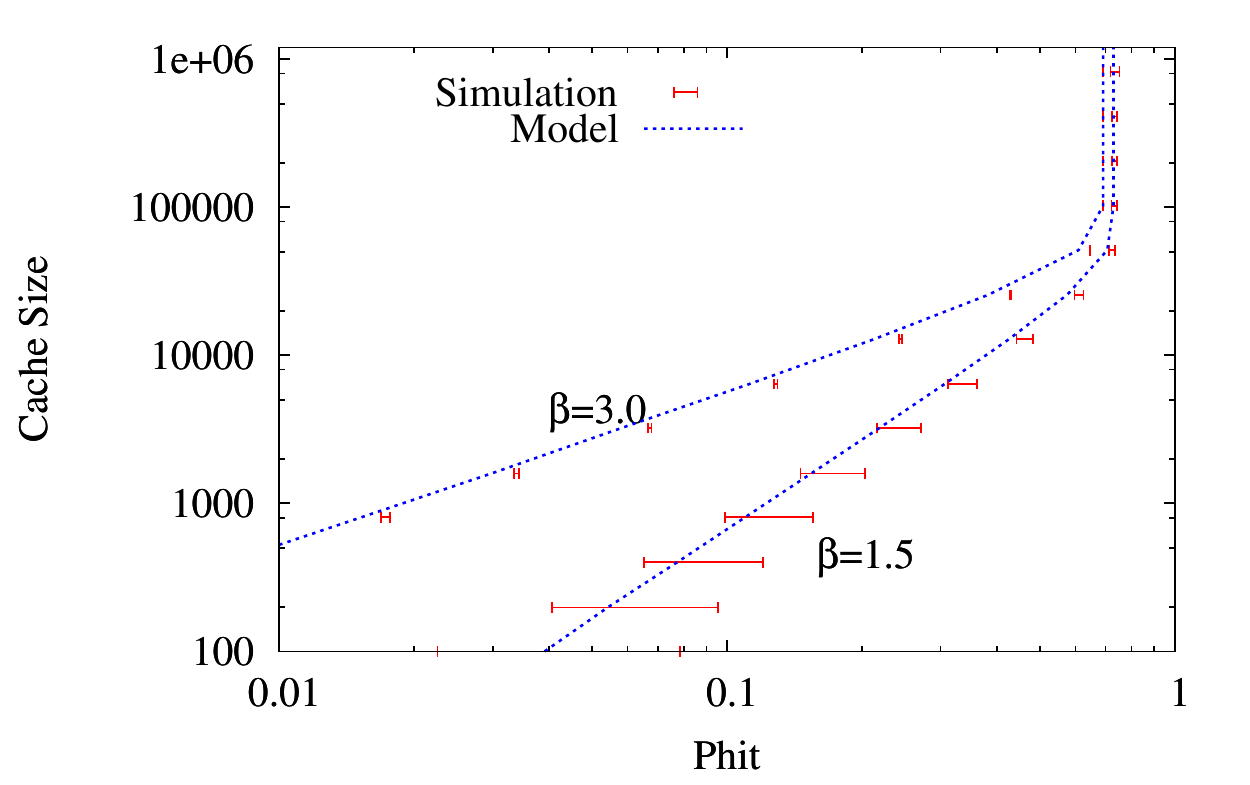}
   \caption{The impact of two different values of Pareto exponent
   $\beta$ (and the corresponding Zipf's $\alpha$).}
    \label{fig:cachesize_vs_phit_uniform_beta_log_mod_sim}
  \end{subfigure}
  \hspace*{0.1cm}
  \begin{subfigure}[t]{0.32\textwidth}
    \centering
    \includegraphics[width=1.05\linewidth]{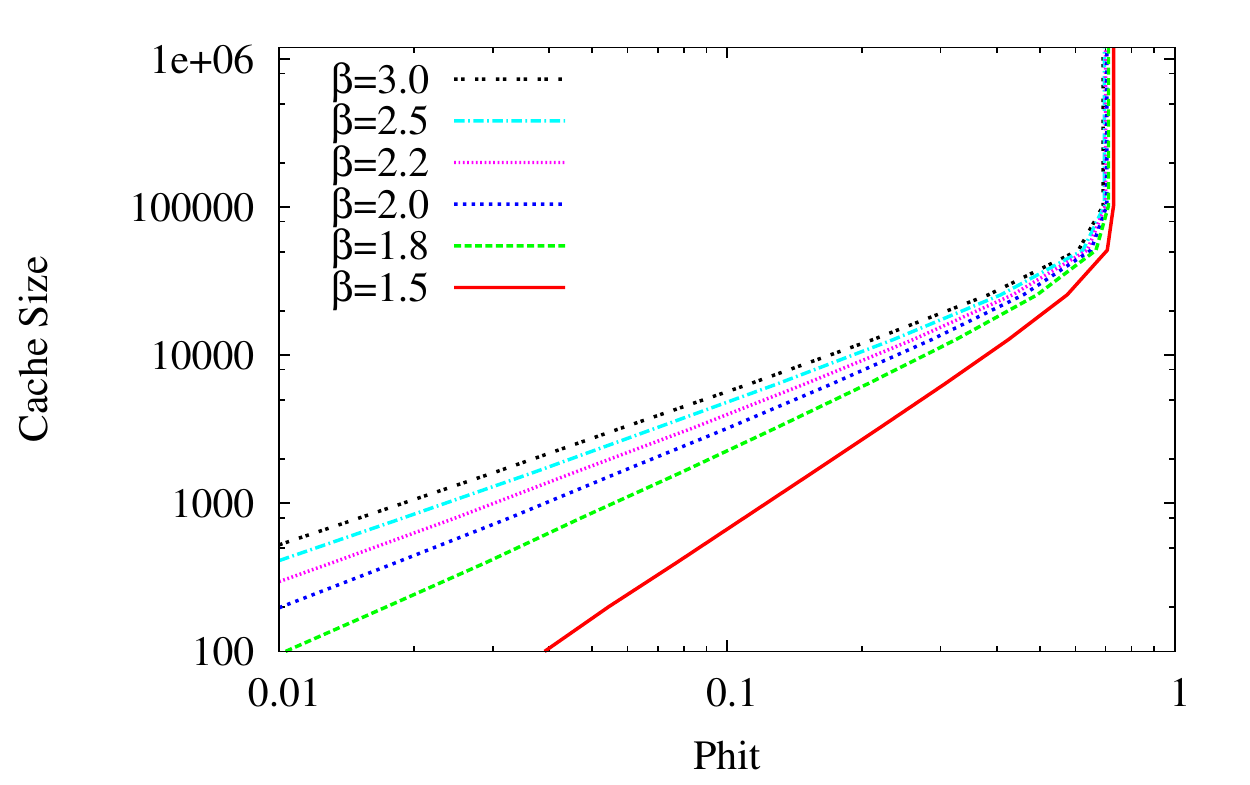}
    \caption{The model results for various values of the Pareto exponent
     $\beta$.}
    \label{fig:cachesize_vs_phit_uniform_beta_log_mod}

  \end{subfigure}
  \vspace*{-0.1cm}
  \caption{Cache performance when varying the average life-span
   $L$, in the case of a Pareto distribution (with $\beta=3$) of requests' volume
   (Fig.~\ref{fig:cachesize_vs_phit_uniform_mod-sim_log_approx}); and while
   varying the Pareto exponent $\beta$, in the case of average content life-span $L=7$
   days (Figs.~\ref{fig:cachesize_vs_phit_uniform_beta_log_mod_sim}
   and~\ref{fig:cachesize_vs_phit_uniform_beta_log_mod}).}
  \vspace*{-0.3cm}
\end{figure*}
To make the previous argument more rigorous, we introduce a {\em virtual
 time} function represented by a generic increasing and continuously
 differentiable function $w(t)$, whose first derivative $w'(t)$ is
 proportional to the desired instantaneous aggregate request rate at
 time $t$.  Function $w(t)$ satisfies the following additional
 properties: $w(0)=0$, $\lim_{t\to \infty} {w(t)}/{t}=1$.  Then we can 
 specify a generalized SNM whereby all temporal dynamics are defined
 over the virtual time $w(t)$ (which replaces the original real time
 $t$).  In particular, the starting time and the popularity
 profile of each individual content are transformed according
 to: $\tau_m \to w(\tau_m)$ and $\lambda(t-\tau)\to \lambda(w(t)-
 w(\tau)) w'(t)$.  In doing so, we obtain the desired effect of
 applying the amplitude modulation $w'(t)$ to the original
 process. Indeed, by construction, the average aggregate instantaneous
 request rate at time $t$ becomes:\mmm\mmm
$$
\lim_{\Delta t \to 0} \frac{\mathbb{E} [{\sum_{m} \int_t^{t+\Delta t }  V_m\lambda (w(t) - w(\tau_m)) w'(t)  d t  }]}{\Delta t}= \gamma \mathbb{E}[V] w'(t)
$$
We can prove the following:
\begin{teorema}
Under Che's approximation, cache performance is invariant under
the transformation $t \to w(t)$.
\end{teorema}
\begin{IEEEproof}
We follow exactly the same lines as in the proof of
Theorem~\ref{theo:1}. In particular, the expression for
$p_{\text{hit}}$ can be obtained from \eqref{eq:phit} and
\eqref{eq:c2} by substituting $\tau$ with $w(\tau)$, $\theta$ with
$w(\theta)$, $d\tau \to w'(\tau)d\tau$, $d\theta \to
w'(\theta)d\theta$. Then the invariance property of $p_{\text{hit}}$
derives from the standard change of variable rule inside the
integrals.\end{IEEEproof}
\begin{remark}\label{remark:5}
Day-night fluctuations in the arrival rate of requests have no
impact on cache performance, so long as such fluctuations are roughly
synchronized at the ingress points of the cache network, i.e., when
users reside in the same time-zone (or in few adjacent time-zones).
Diurnal variations can be important in cache systems covering
several time zones. In this paper we do not investigate the effects of 
different time-zones, because they are not observed in our data-set, 
and therefore cannot be validated with any reasonable level of confidence.  
Nonetheless, if needed, our SNM traffic (and the relative analysis) 
can be easily extended  to incorporate ``out-of-phase'' fluctuations at 
different  ingress points.
\end{remark}\vspace*{-0.3cm}

\vspace*{-0.2cm}
\section{Numerical evaluation}\label{sec:validation}

The goal of this section is two-fold. First, we assess the accuracy of
the analytical model for the cache hit probability, described
in Sec.~\ref{sec:modelling}. Second, we exploit the
insights gained from the model to better understand the performance of
caching systems in the presence of temporal and geographical locality,
showing that our analysis can be useful for system design
and optimization.
We compare the results obtained by the model against Monte-Carlo
simulations of LRU, using the same synthetic SNM traffic considered in
the analysis.  By so doing, we are able to decouple the
errors arising from modeling approximations, from those that derive
from  a non-perfect match between experimental (trace-driven) and synthetic traffic
patterns, which have been discussed before (see Fig.~\ref{fig:phit_vs_cachesize_fertility_flip_trace_04} 
in Sec.~\ref{subsec:valid_single}, and Fig.~\ref{fig:p-pul-pdf}  in Sec.~\ref{subsec:validate_multi}).
Moreover, simulating LRU under the SNM traffic
model enables us to explore a much wider range of scenarios than are
present in our data set, and provides us with fundamental insights
into the impact of the various traffic parameters.

\vspace*{-0.4cm}
\subsection{Single-cache, single-class scenario}\label{sec:singleclass}
We start by considering the basic case of a single cache fed by a
single-class SNM traffic. We set the arrival rate of new contents
($\gamma$) to $10,000$ units per day 
 and assume the average
number of requests ($V$) attracted by each content to follow a Pareto
distribution:
\( f_V(v)=\beta V_{\min}^\beta/v^{1+\beta} \), for $v\ge
V_{\min}$\footnote{ Note that the second moment of $V$ is finite for
  $\beta>2$}.
The choice of a Pareto distribution for $V$ is justified by two factors.
First, previous works have shown that the aggregate requests attracted
by many types of contents (including popular movies or user-generated
videos) over long time periods are well described by the Zipf's
law~\cite{Roberts-ITC}.
Second, a Zipf-like distribution is obtained when a large number of individual
content request volumes are generated independently
according to a Pareto distribution with exponent $\beta$.
For the experiments presented in this section, we fix the average
number of requests for each content to $\mathbb{E}[V]=3$. Since the
shape of the popularity profile has been shown to have a negligible
impact on the resulting cache performance (see
Sec.~\ref{sec:traffic_model} and~\ref{sec:modelling}), unless otherwise specified 
we assume a
uniform popularity profile, with average life-span $L$.
Finally, for the results obtained by simulation, we show the error bars
corresponding to $95\%$ confidence intervals.
 
Fig.~\ref{fig:cachesize_vs_phit_uniform_mod-sim_log_approx}
shows the required cache size to achieve a given hit probability,
for different values of $L$. We observe an almost perfect match
between simulation results (the horizontal error-bars appear as
points) and the model prediction from~\equaref{phit} (dotted lines).
We find that our small-cache approximation~\equaref{psmall} (solid
line) is very accurate for a wide range of values of $p_{\text{hit}}$.
As expected, cache performance is deeply impacted by the average life-span of contents
($L$): as suggested by the closed-form approximation
\equaref{psmall}, for a given cache size, the hit probability is
roughly inversely proportional to the average life-span ($L$).  

To investigate the impact of the distribution of the number of
requests attracted by contents ($V$),
Figs.~\ref{fig:cachesize_vs_phit_uniform_beta_log_mod_sim}
and~\ref{fig:cachesize_vs_phit_uniform_beta_log_mod} show the results
obtained when varying the value of the Pareto exponent $\beta$.
Comparing the analytical prediction~\equaref{phit} 
against simulations for two extreme values of $\beta$, in
Fig.~\ref{fig:cachesize_vs_phit_uniform_beta_log_mod_sim}, we observe
that the model is very accurate.
Fig.~\ref{fig:cachesize_vs_phit_uniform_beta_log_mod} reports the
results for a wider range of $\beta$; for the sake of clarity, here we
omit the simulation results, since we observed a strong
agreement between model and simulation results in all cases.

As expected in general the  distribution of the number of
requests attracted by contents ($V$),  may play an important rule on the cache performance (i.e., the cache size required to achieve a given hit probability); 
the  performance  of the caching system benefits  from  making the popularity distribution less and less skewed (i.e., by decreasing $\beta$). 
Note, however that the {\em  impact on cache performance of the specific $\beta$ is  fairly limited as long as $\beta>2$ (i.e., the variance of the number of
content requests  keeps finite).}  
This is in sharp contrast to results obtained under IRM traffic, where  a small change of the Zipf's exponent has a huge impact on cache performance~\cite{Roberts-ITC}.
As a consequence,  {\em 
 under 
SNM, a precise characterization of popularity distribution parameters
  is not that important to predict cache performance  (as long as the number of
requests attracted by contents  has a finite variance).}

\mmm\mmm\mmm
\subsection{Single-cache, multi-class scenario}\label{sec:multiclass}
We now move on to the case of a single cache fed by a multi-class SNM
traffic,  with the goal of understanding the impact on cache
performance of a mixture of highly heterogeneous contents characterized
by different degrees of temporal locality. This is indeed 
the kind of traffic that we observe in a real network, as 
we found in our data set (see~\autoref{tab:desc-dataset}). 
In particular, we consider the $6$ classes of contents listed
in~\autoref{tab:mix1}, whose parameters have been chosen 
to reasonably match a realistic scenario (see Sec.~\ref{sec:analysis}).  
Class~0 collects unpopular
contents with request volumes smaller than 10.
Classes $1$--$5$ correspond to popular contents having different degree of
temporal locality, with average life-span ($L$) ranging from a few
days (Class~1) to several years (Class~5). The different values for
the average number of requests attracted by contents in these classes
reflect the observations from our traces (see~\autoref{tab:desc-dataset}). 

In order to understand the impact of different traffic mixes, we
consider 3 traffic scenarios in which we vary the proportion of each
class of contents 
(i.e., the probability $\Prob\{W_1=k\}$ that a new content belongs to a given class),
as reported in the last $3$ columns of~\autoref{tab:mix1}. Note that
Class~1 is missing in both \textit{Scenario 1} and \textit{Scenario
  2}, whereas Class~2 is missing only in \textit{Scenario 1}.  For all
scenarios the arrival rate of new contents is set to $\gamma = 10^5$
contents/day.  Finally, an exponential popularity profile is chosen
for all the classes.

\begin{table}[]
\centering
\tiny
\begin{tabular}{|c||c|c|c|c||c|c|c|}
\hline
Class &  ${L}$ (days) & $E[{V}]$ & $V_{\text{max}}$  & $\beta$ & Scen.~1 & Scen.~2 & Scen.~3\\
\hline
 0  & 500    & 1.61  & 10    & 2.5  & 0.85   & 0.85  & 0.85     \cr
 1  & 2      & 83.33   & $\infty$   & 2.5 & 0.00 & 0.00 & 0.01\\
 2  & 7      & 75.00   &  $\infty$  & 2.5  &  0.00     & 0.02    & 0.02    \cr
 3  & 30     & 66.66   & $\infty$   & 2.5  & 0.02    & 0.02  & 0.02   \cr
 4  & 100    & 50.00   &   $\infty$  & 2.5 & 0.02    & 0.02  & 0.02   \cr
 5  & 1000   & 50.00   & $\infty$   & 2.5  & 0.11    & 0.09  & 0.08    \cr

\hline
\end{tabular}
\caption{Content class parameters and their composition for each
  multi-class scenario.}
\label{tab:mix1}
\vspace*{-0.5cm}
\end{table}

\begin{figure}[t!]
  \begin{minipage}[t]{0.48\columnwidth}
    \hspace*{-0.6cm}
    \includegraphics[width=1.15\linewidth]{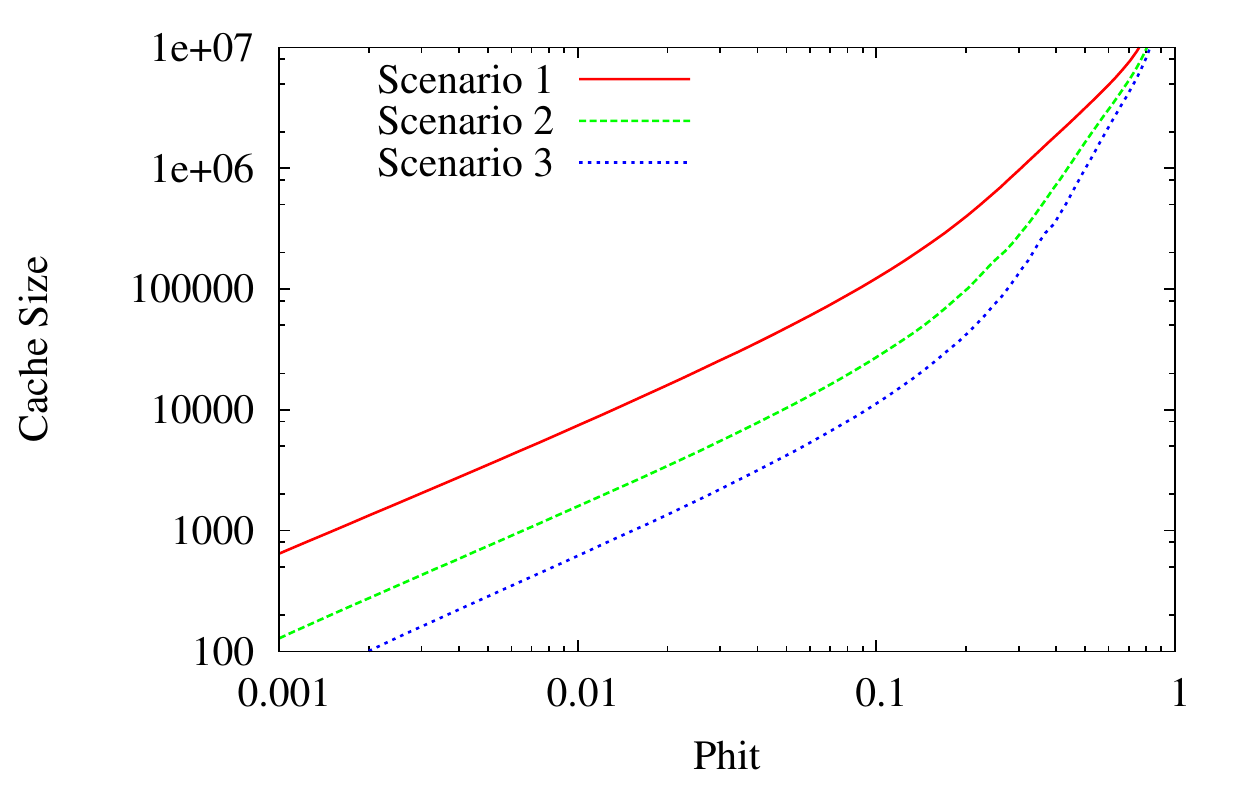}
    \caption{Cache performance for different traffic scenarios.}
    \label{fig:cachesize_vs_phit_multiclass-scenarios_exp}
  \end{minipage}
  \hspace*{0.1cm}
  \begin{minipage}[t]{0.48\columnwidth}
    \hspace*{-0.3cm}
    \includegraphics[width=1.15\textwidth]{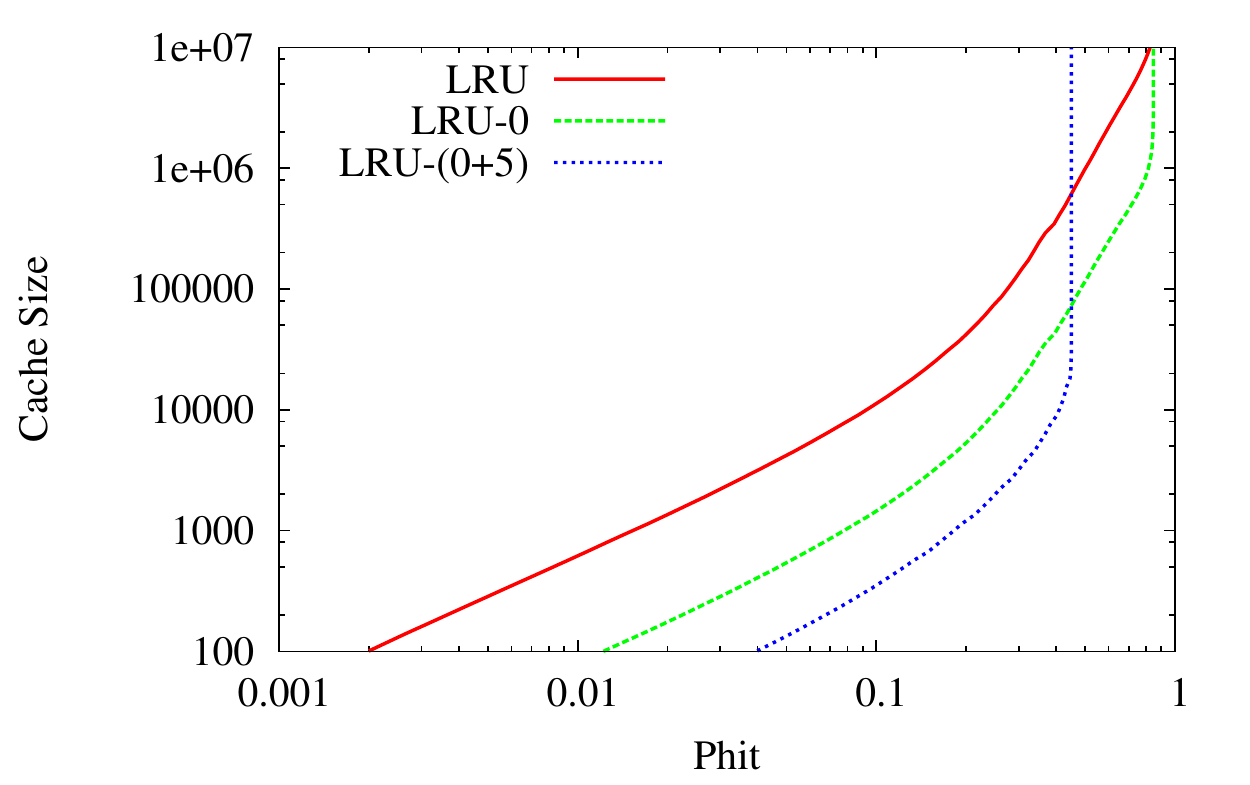}
    \caption{Cache performance when; i) no filtering, ii)
          filtering Class \textit{0}, iii) filtering Classes \textit{0}
          and \textit{5}.}
    \label{fig:cachesize_vs_phit_multiclass-filter_exp}
  \end{minipage}
  \vspace*{-0.8cm}
\end{figure}

Fig.~\ref{fig:cachesize_vs_phit_multiclass-scenarios_exp} reports the
cache performance under the three considered scenarios, as
  obtained  by (\ref{phitformulamulticlass}).
We observe that {\em the presence of just a small fraction of highly
cacheable contents} (in \textit{Scenario 3} only 3\% of the contents
belong to either Class \textit{1} or \textit{2}) {\em has a huge
beneficial impact on the hit probability, especially with small
caches.}  Even for medium-size caches the gain is very significant: for
example, in the case of $C = 10,000$, the hit probability goes from 5\%
(\textit{Scenario 1}) to about 20\% (\textit{Scenario 3}).

Previous results suggest that contents characterized by high temporal
locality, although few in number, do play the major role in the
resulting hit probability.  This fact also suggests that, {\em  when the cache
size is limited, it may be convenient to devote the entire cache space
only to highly cacheable contents (i.e., contents with large volumes
and significant temporal locality), and to forbid other contents from
entering the cache.} This strategy minimizes the
probability of evicting from the cache contents with a high
temporal locality in their request pattern, to 
let room to an unpopular content which will likely not be requested
again while being cached (hence storing this content in the cache
is useless).
To check the extent to which this assertion is valid, 
we modified the classical LRU caching strategy (and the corresponding analysis) such that contents
belonging to specified classes 
are never cached (notice that, by so doing, all requests for filtered
contents deterministically produce a miss).
The extension of the analysis to compute the resulting hit probability on
this LRU variant is rather straightforward, hence we omit the details here.
Under \textit{Scenario 3},
  Fig~\ref{fig:cachesize_vs_phit_multiclass-filter_exp} compares the
  performance of LRU against the performance of LRU-0 and LRU-(0+5),
  which  do not cache contents of class
  \textit{0} and of both classes \textit{0} and \textit{5}, respectively. 
  Observe that, {\em when the cache
size is limited, a significant performance improvement is achieved by
filtering out contents that are either unpopular (class \textit{0}) or
popular but long-lived (class \textit{5}). } For example, the adoption
of LRU-(0+5) leads to a reduction of more than one order of magnitude
in the cache size needed to achieve $p_{\text{hit}}=0.1$, with
respect to LRU. Finally, as expected, filtering out contents when the
cache size increases must at some point become deleterious,
since filtered contents lead to a miss in the cache. This is confirmed by
the intersection between the curves in
Fig~\ref{fig:cachesize_vs_phit_multiclass-filter_exp}.

The practical implementation of filters to detect unpopular/long lived
contents raises issues that go beyond the scope of this paper. Here we
limit ourselves to mentioning that content classification can be
accomplished either by exploiting a-priori information about the
content, such as the category, the producer etc., or by employing
blind online techniques to infer the instantaneous request rate
subject to the history of requests~\cite{jel08}.

\vspace*{-4mm}
\subsection{Multi-cache, single-class scenario} \label{sect:result-multicache}
We now consider a simple cache network with a tree structure. In
addition to assessing the accuracy of the improved approximation
described in Appendix~\ref{sec:nets}, this scenario permits us to
understand the impact of geographical locality on cache performance.
In more detail, we consider a two-layer cache network composed by $8$
leaves and one root (plus an additional repository above the root).
As in the standard Edge CDN
architectures~\cite{Nygren:2010,Huang:2013}, content requests arrive
at the leaves, and misses are forwarded to the root.  Here, we focus
on two extreme traffic scenarios: i) an {\em unlocalized} scenario, in
which content requests are equally likely to arrive at any of the
leaves (independently at random), and ii) a {\em fully localized}
scenario, in which each leaf receives the requests for just a subset
of the entire catalog (i.e., each newly introduced content is
statically assigned to a distinct leaf, which will receive all its
associated requests).
For both scenarios, we consider a single-class SNM with the following
parameters: requests volumes $V$ are Pareto-distributed with $\mathbb{E}[V]=3.0$ and
$\beta=2.5$; the popularity profile is exponential with average life-span $L=7$~days;
the aggregate arrival rate of new contents ($\gamma$) is set to
$10,000$ contents per day.

Being conscious that the extreme scenarios proposed above are
oversimplified and may look somehow artificial, we emphasize that our
goal here is to understand the potential impact of geographical locality on
the overall performance of a caching system. Hence, by considering the
two extreme cases above, we can evaluate the whole range of possible
behavior of the system under intermediate (more realistic) traffic
patterns.  

\begin{figure}[t!]
  \begin{subfigure}[t]{0.48\columnwidth}
    \hspace*{-0.6cm}
    \includegraphics[width=1.15\linewidth]{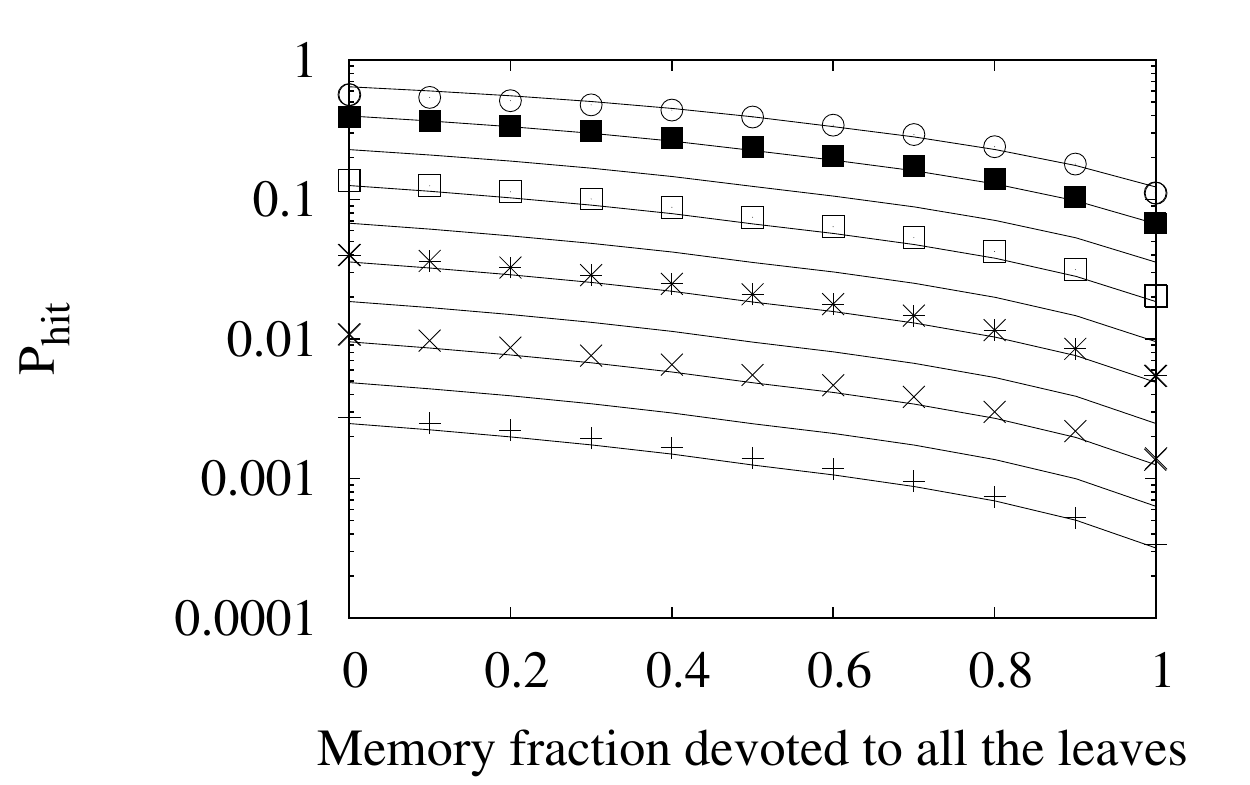}
    \caption{Unlocalized scenario.}
    \label{fig:delocal_multicache}
  \end{subfigure}
  \hspace*{0.1cm}
  \begin{subfigure}[t]{0.48\columnwidth}
    \hspace*{-0.4cm}
    \includegraphics[width=1.15\textwidth]{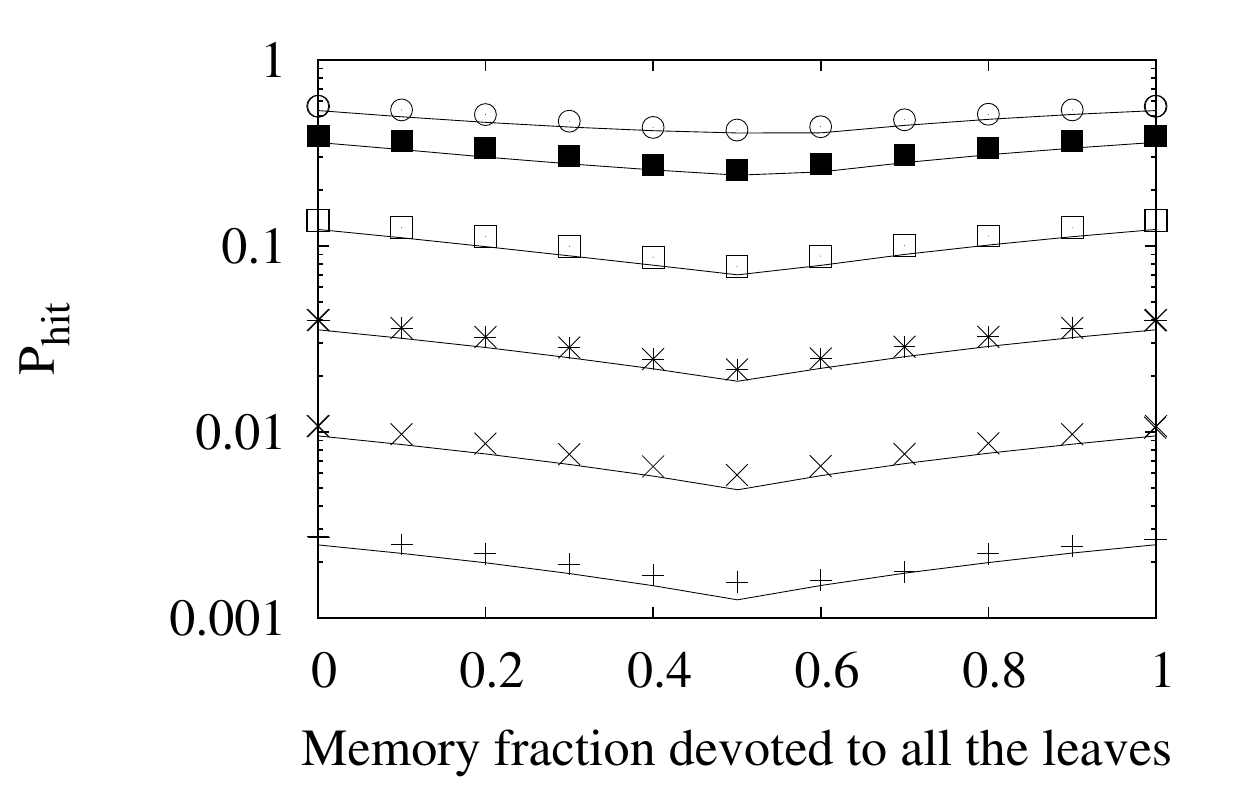}
    \caption{Localized scenario.}
    \label{fig:local_multicache}
  \end{subfigure}
 \vspace*{-0.1cm}
  \caption{Hit probability for different cache size under different traffic localization scenarios. Lines refer to the analytical model, points
  to the simulation results. Caches sizes are (from
  bottom to top) $100$, $400$, $1600$, $6400$, $25600$, $51200$ contents.}
  \label{fig:de-local_multicache}
 \vspace*{-0.6cm}
\end{figure}

Fig.~\ref{fig:de-local_multicache} reports the global hit probability
(i.e.\ either at any leaf or at the root cache), for the {\em
  unlocalized} (Fig.~\ref{fig:delocal_multicache}) and {\em fully
  localized} (Fig.~\ref{fig:local_multicache}) scenario, as function
of the fraction of total storage capacity assigned to the leaves
(i.e., we assume that the total capacity of all caches is kept
constant).  We show both the results obtained analytically with the
improved approximation explained in Appendix~\ref{sec:nets}, and
simulation results obtained under the same SNM. In both scenarios, the
analytical predictions (lines) match very well simulation results
(marks).
Beyond proving the accuracy of the model, some interesting insights at
system level can be obtained from the plots in Fig.~\ref{fig:de-local_multicache}.
{\em When no geographical locality is present} (Fig.~\ref{fig:delocal_multicache}),
{\em the maximum hit probability is achieved when the whole storage
capacity is located in a single cache (the root).} This can however
result in longer access delays for the users.
{\em Instead, when traffic is strongly localized, } (Fig.~\ref{fig:local_multicache}), 
{\em by increasing the cache size of the
leaves (up to the point at which all storage capacity is assigned to
the leaves) we jointly maximize the cache hit probability while
reducing the access delay.} Interestingly, in this case the same
maximum hit probability is also achieved by putting all storage in the
root, although this would be detrimental in terms of delay.

At last we emphasize that  geographical locality plays a similar role also  
under a multi-class scenario, for which do not report results  
due to space constraints.

\vspace*{-0.2cm}
\section{Conclusions}\label{sec:conclusion}
The Shot Noise Model provides a simple, flexible and accurate approach to 
describing the temporal and geographical locality found in Video-on-Demand
traffic, allowing us also to develop accurate analytical models
of cache performance. From the point of view of system design,
our main findings are: 
i) cache performance can significantly benefit from the presence of even 
a relatively small portion of highly cacheable (popular and local) contents (especially when 
caches are small); 
ii) geographical locality plays also an important role in the dimensioning of  distributed caching systems and should  
not be neglected; 
iii) the overall impact on cache performance of the 
distribution of the number of requests attracted by the contents 
(and the corresponding rank distribution) is significantly mitigated by 
temporal locality with respect to traditional stationary models (e.g., IRM);
iv) especially when caches are small, performance can be 
significantly improved by restricting access into the cache only to 
contents which are highly cacheable. This can be obtained 
either exploiting a priori information about the contents'
nature and popularity profiles, or by measuring the content 
instantaneous popularity.



\vspace{-4mm}
\appendices
\linespread{0.92}
\section{Proof of Theorem 1}\label{app:theo1}
\begin{IEEEproof}
By adopting Che's approximation, 
we assume that the cache eviction $T_C$ time is
  constant and independent of the considered content.  We
  express the probability of finding a given content $m$ in the
  cache at time $t$, conditionally  on its starting time ($\tau_m$) and
average  request volume ($V_m$), as:\mmm
\begin{equation}\label{eq:ph2}
  p_{\text{in}}(t\mid \tau_m,V_m) = 1- e^{-V_m \int_ {t-T_C}^{t}\lambda(\theta-\tau_m)d\theta}\mmm
\end{equation}
since, under LRU, the considered content is found in the cache at time $t$, iff
it attracted at least one request in the time interval \mbox{$[t-T_C,t]$}.
Unconditioning (\ref{eq:ph2}) with respect to $V_m$, and recalling that $V_m$ are i.i.d.\ as $V$, we obtain:\mmm\mmm
\begin{multline*}
  p_{\text{in}}(t\mid \tau_m)=\mathbb{E}_{V}\left[ 1-  e^{-V \int_ {t-T_C}^{t}\lambda(\theta-\tau_m)d\theta} \right]= \\
 \vspace{-3mm} 1- {\phi_{V} \left(-\int_{t-T_C}^{t} \lambda(\theta-\tau_m)d\theta
    \right)}\mmm
\end{multline*}

To evaluate the probability of finding content $m$ in the cache
at time $t$, we uncondition the above expression with respect to $\tau_m$.
To do this, recall that in a Poisson process, conditionally over the number of
  points falling in the interval $[0,t)$,
     each point is uniformly distributed
    over the considered interval, independently of other points. 
Hence, the distribution of $\tau_m$ is uniform in the interval $[0,t)$, and we obtain:\mmm\mmm
\begin{equation}\label{eq:pin}
  p_{\text{in}}(t)=\frac{1}{t}\int_{0}^t
  1- {\phi_V \left(-\int_ {t-T_C}^{t} \lambda(\theta-\tau)d\theta \right)} d \tau  \mmm
\end{equation}

Now, as in the standard IRM, for a sufficiently large $t$, we can
assume that the cache is completely filled with contents introduced
before $t$, and the number of contents in the cache is exactly equal
to its size.  Therefore we can write:\mmm\mmm
\[
C= \sum_m [\Uno_{\{\text{content } m\text{ in cache at time }t\mid \tau_m\le t\}}
\Uno_{\tau_m\le t}]\mmm
\] 
where the
sum extends over all contents in the infinite content catalogue.
Averaging both terms we obtain:\mmm\mmm
\begin{equation}\label{eq:c}
  C= \sum_m \mathbb{E} [ \Uno_{\{\text{$m$ in cache at $t$}\mid
      \tau_m\le t\}} \Uno_{\tau_m\le t}]= p_{\text{in}}(t) \sum_m
  \mathbb{E}[\Uno_{\tau_m\le t}]\mmm
\end{equation}
Recalling that the average rate at which new contents are introduced is
$\gamma$, by combining~\eqref{eq:pin} with~\eqref{eq:c}, we 
can express the size of the cache $C$ as:\mmm\mmm
\begin{multline}\label{eq:cg}
\hspace{-5mm}
  C = \left(\sum_m \frac{\mathbb{E}[\Uno_{\tau_m<t}]}{t}\right)
  \int_{0}^t 1- \phi_V\left(-\int_ {t-T_C}^{t} \lambda(\theta-\tau)d\theta \right) d \tau  
  = \\ 
  \gamma \int_{0}^t 1- \phi_V \left(-\int_ {t-T_C}^{t} \lambda(\theta-\tau)d\theta \right) d \tau \mmm
\end{multline}
Eq.~\eqref{eq:cg}, proving~\eqref{eq:c2}, must be solved
(numerically) to evaluate the eviction time ($T_C$) for a given cache size
$C$. Furthermore, \eqref{eq:cg} provides an
  interesting insight into the cache behavior: {\em for a given  eviction time $T_C$, the cache size $C$ is proportional to the rate at
  which new contents are introduced ($\gamma$)}.

Now we turn our attention to the hit probability.
Assume that a request $R_t$ 
 arrives at the cache at time $t$ for content $m$ of parameters $(\tau_m,V_m)$. 
By definition, $R_t$ generates a cache hit iff the content is found in
the cache. Therefore, as consequence of the Lack of Anticipation
(LAA) property~\cite{wolf1989stochastic} of the request process for content
$m$, the hit probability experienced by request $R_t$ is:\mmm\mmm
\[
 p_{\text{hit}}(t\mid \tau_m,V_m)= p_{in}(t\mid \tau_m,V_m)\mmm
\]

Now, when unconditioning $p_{\text{hit}}(t\mid \tau_m,V_m)$ with respect to
$(\tau_m, V_m)$, we have to carefully account for the fact  that contents are not uniformly requested.
Observe that the instantaneous request at which cache requests  rate for a specific content $m$ is given
$V_m\lambda(t-\tau_m)$.   
Thus,
we can interpret:\mmm
\[ V_m\lambda(t-\tau_m) \cdot p_{\text{hit}}(t\mid \tau_m,V_m) \mmm\] 
 as the  hit rate generated by content $m$.
Summing up all contents, we can express $ p_{\text{hit}}(t)$ as the ratio between average global cache hit-rate and average global request rate. It turns out that:\mmm\mmm\mmm
\[
 p_{\text{hit}}(t)= \frac{\mathbb{E}[\sum_m  V_m\lambda(t-\tau_m) \cdot p_{\text{hit}}(t\mid \tau_m,V_m)]}{\mathbb{E}[\sum_m  V_m\lambda(t-\tau_m) ] }\mmm
\]
Recalling~\eqref{eq:ph2} we have for $t\ge T_C$:\mmm\mmm\mmm\mmm
\begin{multline*}
  p_{\text{hit}}(t)=\\
 \frac{\gamma \int_0^t \mathbb{E}_{V}  \left[V
      \lambda(t-\tau)\left( 1- e^{-V \int_
        {t-T_C}^{t}\lambda(\theta-\tau)d\theta}\right)\right]
  d\tau}{\gamma \mathbb{E}[V] \int_0^t \lambda(t-\sigma) d\sigma }=\\
  \int_{0}^t \lambda(t - \tau) \left(1-\frac{\phi_V' \left(-\int_
        {t-T_C}^{t} \lambda(\theta-\tau)d\theta
      \right)}{\mathbb{E}[V]\int_0^t\lambda(t-\sigma)d\sigma} \right)d \tau\mmm\mmm
\end{multline*} 
Substituting $\alpha=t-\tau$, $\beta=t-\theta$, $\zeta=t-\sigma$  we get:\mmm\mmm
\begin{equation}\label{eq:ph1}
  p_{\text{hit}}(t)=
  \int_{0}^t \lambda(\alpha) \left(1-\frac{\phi_V'   \left(-\int_ {0}^{T_C} \lambda(\alpha-\beta)d\beta 
      \right) } {\mathbb{E}[V]\int_0^t\lambda(\zeta)d\zeta} \right) d \alpha \mmm
\end{equation} 
Thanks to the integrability property of $\lambda(t)$, (\ref{eq:phit})
is obtained by letting $t \to \infty$ in~\eqref{eq:ph1}.
\end{IEEEproof}

\vspace*{-4mm}
\section{LRU in cache networks}\label{sec:nets}

We now show how our analysis of LRU policy under SNM can be
extended to the case of a network of caches, whereby misses are forwarded
to other caches, according to pre-established routes,
possibly ending up at a repository storing
the entire catalogue.
In doing so, we will propose an improved approximation {with
respect to the one  proposed in~\cite{kurose2010}}, which is based on the simplifying
  assumption that the arrival process of requests for a given content
  at any cache is Poisson.
For simplicity, we will consider only networks implementing the
so-called leave-copy-everywhere replication strategy,
according to which a copy of a requested content is inserted in all
caches traversed by the request. Moreover, we will restrict ourselves
to the case of tree-like topologies~\footnote{The extension of our
  analysis to general mesh networks can be carried out in a similar
  way as proposed in~\cite{kurose2010} under the Poisson
  approximation.  This extension is conceptually very simple, but
  requires a global multi-variable fixed point procedure to solve the
  entire system.}.  Requests arrive initially at the leaves of
the tree. Whenever a content is not found at a leave, it is forwarded
to the parent node. We assume that there exists a repository storing
the entire catalogue above the root of the tree.

\mmm\mmm
\subsection{The Poisson approximation}
For any cache $c$ in the network, we denote by $\mathcal{C}(c)$ the
set of children of $c$ (caches) in the tree.  Let $\mathcal{F}$ be the
set of caches corresponding to the leaves of the tree. We denote by
$\mathcal{F}(c)$ the subset of $\mathcal{F}$ corresponding to the
descendants of $c$ in the tree.
According to the model proposed in
Sec.~\ref{subsec:snm_multi}, at time $t$, the arrival rate of requests for content
$m$ at leaf node $f\in \mathcal{F}$ is given by:\mmm\mmm
\[
\lambda^{(f)}(t\mid V_m, \tau_m )= V_{m,f} \lambda_m(t- \tau_m)  = V_m p_{m,f} \lambda_m(t- \tau_m)
\mmm
\]
where $V_m$ is the total request volume produced by content $m$ and
$\tau_m$ is the time at which content $m$ is introduced into the
catalogue. Recall also that $p_{m,f} $ represents the probability that
a request for content $m$ enters the network at cache $f$.  By
construction, $\sum_{f\in \mathcal{F}} p_{m,f} = 1$.  We observe that
each leaf can be independently analyzed using  \eqref{eq:phit}  and \eqref{eq:c2}.

Consider now a non-leaf node $c$; the intensity of the arrival process
of requests for content $m$ at $c$ at time $t$ is given by:\mmm\mmm
\[\hspace{-3mm}
 \lambda_m^{(c)}(t\mid V_m, \tau_m)  = \!\!\!\! 
 \sum_{c'\in \mathcal{C}(c)} \lambda_m^{(c')}(t\mid V_m, \tau_m)(1 -\Uno_{\{m\in c' \}}(t\mid V_m, \tau_m))
 \mmm
\]
where $\Uno_{\{m\in c' \}}(t\mid V_m, \tau_m)$ is the indicating
function corresponding to the event $\{ m\in c' \text{ at time $t$ }
\mid V_m, \tau_m \}$.  As such, $\lambda^{(c)}_m(t\mid V_m, \tau_m)$
turns out to be a stochastic variable, because its value dependent on
the state of caches in $\mathcal{C}(c)$.  This implies that the
arrival process of requests at cache $c$ is not an inhomogeneous
Poisson process.  The expectation of $\lambda^{c}_m(t\mid
V_m, \tau_m)$ can be computed as:\mmm\mmm
 \begin{multline}
 \lambdahat_m^{(c)}(t\mid V_m, \tau_m)  =\mathbb{E}[ \lambda_m^{(c)}(t\mid V_m, \tau_m) ] = \\
= \sum_{c'\in \mathcal{C}(c)} \mathbb{E}[ \lambda_m^{(c')}(t\mid V_m, \tau_m) ] 
 (1 - p^{(c')}_{\text{in}} (t\mid V_m, \tau_m) ) 
\label{eq:b16}\mmm\mmm
\end{multline}
where $p^{(c')}_{\text{in}}(t \mid V_m, \tau_m)$ is the probability
that content $m$ (with attributes $(\tau_m, V_m)$) is cached in
$c'$ at time $t$.

The standard approximation to (\ref{eq:b16}) would be to replace
$\lambda^{c}_m(t \mid V_m, \tau_m)$ with its expectation
$\lambdahat_m^{(c)}(t\mid V_m, \tau_m)$ at all nodes which are not
leaves, i.e., to approximate the arriving process of requests for
content $m$ at $c$, with an inhomogeneous Poisson process whose
instantaneous intensity at time $t$ is equal to the average
intensity of the actual process at time $t$, and then independently
solve each cache in isolation using the single-cache analysis. 

At last, approaches that go beyond the
Poisson approximation have been recently proposed in~\cite{nain}.  These approaches attempt to better
characterize the cache miss stream. However, they rely
heavily on the assumption that the request arrival process for a given
content at a cache is a stationary (renewal) process, and are
therefore not applicable to our case.
\mmm\mmm
\subsection{Improved approximation} 
Our basic idea is to approximate the
correlation between the states of neighboring caches, which is totally
neglected under the Poisson approximation. 
This can be done by distinguishing between the hit
  probability ($p_{\text{hit}}^{(c)}(t \mid V_m, \tau_m)$) and
  the probability of finding a given content $m$ in the cache at time $t$
  ($p_{\text{in}}^{(c)}(t \mid V_m, \tau_m)$). Note that, while the hit
  probability is implicitly conditioned to the event that a request
  for $m$ arrives at cache $c$ at time $t$, the probability of the
  content being present in the cache is not. By virtue of the lack of
anticipation property \cite{wolf1989stochastic}, the above two
probabilities would be equal if the arrival process was an
inhomogeneous Poisson process. However, the arrival process at
non-leaf nodes is not Poisson,
hence the above two probabilities can be different at non-leaf nodes.

{To approximately
  evaluate $p_{\text{hit}}^{(c)}(t \mid V_m, \tau_m)$, 
 we focus on a request
  for content $m$ arriving at cache $c$ at time $t$ from a given child
  node $c'$.  We observe that such a request can arrive at time $t$ at
  cache $c$, only if content $m$ is not stored in cache $c'$ at
  $t^-$. This implies that no request for $m$ can have arrived to $c'$
  in the interval $[t- T^{(c')}_C, t]$ (otherwise content $m$ would be
  stored in cache $c'$ at time $t^-$). Therefore, a fortiori, no
  request for $m$, coming from $c'$, can have arrived at cache $c$ in
  the same interval.  Indeed, content $m$ is found in cache $c$ at time
  $t$ by a request arriving from $c'$ if and only if either (i) at
  least one request arrived at cache $c$ within the interval
  $[t-T^{(c)}_C, t-T^{(c')}_C]$ from any child node, including $c'$
  (this case is considered provided that $T^{(c)}_C > T^{(c')}_C$); or
  (ii) at least one request arrived at cache $c$ within the interval
  $[t-T^{(c')}_C, t]$ from $c'' \neq c'$, i.e., from caches different
  from $c'$ (since we know that no request can arrive at $c$ from $c'$
  during this interval)}.

During both intervals considered above, the arrival process of
requests at cache $c$ from any child node $c'$ is not Poisson (but depends
on the unknown state of the child node), and lacking a better
approach, we resort to approximating it by a Poisson process
with the expected intensity.
By so doing we can compute the conditioned hit
  probability $p_{\text{hit}} (t \mid V_m, \tau_m, c')$ for requests
  coming from $c'$ as:\mmm\mmm\mmm\mmm
\begin{multline*}
p_{\text{hit}} (t \mid V_m, \tau_m, c') \approx 1 -   
e^{-  \int_{t-T^{(c) }_C}^{t-\min(T^{(c')}_C, T^{(c)}_C)}  \lambdahat_m^{(c)} (\tau \mid V_m, \tau_m) \diff \tau} \cdot \\
\prod_{c''\in \mathcal{C}(c)\setminus c' } e^{-  \int_{t-  \min(T^{(c')}_C, T^{(c)}_C)  }^t 
\lambdahat_m^{(c'')}(\tau \mid V_m, \tau_m)\left(1- p_{\text{in}}^{(c'')}(\tau \mid V_m, \tau_m)\right) \diff \tau }\mmm\mmm
\end{multline*}
\vspace{-0mm}
Unconditioning with respect to $c'$ (i.e., by properly taking into
account the fraction of requests for $m$ arriving at $c$ at time $t$
from each child), we obtain an approximate expression for the overall
hit probability of content $m$ at cache $c$ at time $t$ (we omit the
details of this unconditioning). 
The above reasoning cannot be applied to the computation of
$p_{\text{in}}^{(c)}(t \mid V_m, \tau_m)$, for which we resort to the
standard Poisson approximation. 

\end{sloppypar}
\end{document}